\documentclass[twocolumn,prc,aps,showpacs,superscriptaddress, 10pt]{revtex4-1}
\usepackage{graphicx,graphics}
\graphicspath{grafics/}
\usepackage{amsmath}
\usepackage{amssymb}
\usepackage{bm}
\usepackage{epstopdf}
\usepackage{amstext}
\usepackage{amsthm}
\usepackage{amsfonts}
\usepackage{latexsym}
\usepackage{array}
\usepackage{xfrac}
\usepackage{color}
\usepackage{subfigure}
\usepackage{multirow}
\usepackage{fontenc}
\usepackage{tabularx}
\usepackage{tabu}
\usepackage{adjustbox}
\usepackage{dcolumn}
\newcolumntype{d}{D{.}{.}{2.5}}
\newcolumntype{s}{D{.}{.}{1.2}}
\DeclareMathOperator{\arcsinh}{arcsinh}
\usepackage{hyperref}
\hypersetup{colorlinks=true,breaklinks,linkcolor=red,citecolor=blue}

\begin{document}
\title{Higher-order symmetry energy and neutron star core-crust transition with Gogny forces}
\author{C. Gonzalez-Boquera}
\affiliation{Departament de F\'isica Qu\`antica i Astrof\'isica and Institut de Ci\`encies del Cosmos (ICCUB), 
Facultat de F\'isica, Universitat de Barcelona, Mart\'i i Franqu\`es 1, E-08028 Barcelona, Spain}
\author{M. Centelles}
\affiliation{Departament de F\'isica Qu\`antica i Astrof\'isica and Institut de Ci\`encies del Cosmos (ICCUB), 
Facultat de F\'isica, Universitat de Barcelona, Mart\'i i Franqu\`es 1, E-08028 Barcelona, Spain}
\author{X. Vi\~nas}
\affiliation{Departament de F\'isica Qu\`antica i Astrof\'isica and Institut de Ci\`encies del Cosmos (ICCUB), 
Facultat de F\'isica, Universitat de Barcelona, Mart\'i i Franqu\`es 1, E-08028 Barcelona, Spain}
\author{A. Rios}
\affiliation{Department of Physics, Faculty of Engineering and Physical Sciences, University of Surrey, Guildford, Surrey GU2 7XH, United Kingdom}
\date{\today}

\begin{abstract}
\begin{description}
\item[Background] 
An accurate determination of the core-crust transition is necessary in the modeling of neutron stars for astrophysical purposes. The transition is intimately related to the isospin dependence of the nuclear force at low baryon densities.
\item[Purpose] 
To study the symmetry energy and the core-crust transition in neutron stars using the finite-range Gogny nuclear interaction and to examine the deduced crustal thickness and crustal moment of inertia. 
\item[Methods] 
The second-, fourth- and sixth-order coefficients of the Taylor expansion of the energy per particle in powers of the isospin asymmetry are analyzed for Gogny forces.
These coefficients provide information about the departure of the symmetry energy from the widely used parabolic law. 
The neutron star core-crust transition is evaluated by looking at the onset of thermodynamical instability of the liquid core.
The calculation is performed with the exact Gogny EoS (i.e., the Gogny EoS with the full isospin dependence) for the $\beta$-equilibrated matter 
of the core, and also with the Taylor expansion of the Gogny EoS in order to assess the influence of isospin expansions on locating the inner edge of neutron star crusts. 
  \item[Results] 
The properties of the core-crust transition derived from the exact EoS differ from the predictions of the Taylor expansion even when the expansion is carried through sixth order in the isospin asymmetry.
Gogny forces, using the exact EoS, predict the ranges $0.094 \text{ fm}^{-3} \lesssim \rho_t \lesssim 0.118\text{ fm}^{-3}$ for the transition density and $0.339 \text{ MeV fm}^{-3} \lesssim P_t \lesssim 0.665 \text{ MeV fm}^{-3}$ for the transition pressure. The transition densities show an anticorrelation with the slope parameter $L$ of the symmetry energy. The transition pressures are not found to correlate with $L$. Neutron stars obtained with Gogny forces have maximum masses below $1.74M_\odot$ and relatively small moments of inertia. The crustal mass and moment of inertia are evaluated and comparisons are made with the  constraints from observed glitches in pulsars.
\item[Conclusions] 
The finite-range exchange contribution of the nuclear force, and its associated non-trivial isospin dependence, is key in determining the core-crust transition properties. Finite-order isospin expansions do not reproduce the core-crust transition results of the exact EoS. The predictions of the Gogny D1M force for the stellar crust are overall in broad agreement with those obtained using the Skyrme-Lyon EoS.
\end{description}
\end{abstract}

\maketitle

\section{Introduction}

Neutron stars are unique laboratories that provide access to regimes of extreme isospin and density via astrophysical 
observations \cite{Shapiro1983,Glendenning2000}. A wealth of data in single and binary neutron star 
systems is imposing more and more precise limits on nuclear observables, above but also close to saturation density 
\cite{Haensel2007}.  Observations on neutron stars masses from binaries already restrict nuclear models and their 
isospin dependence \cite{Lattimer2012}, and even more stringent constraints will be available from upcoming accurate 
radius measurements with x-ray telescopes \cite{Watts2016}. The mass and radius of a neutron star is directly related 
to the equation-of-state (EoS) of neutron-rich matter \cite{Oertel:2016bki}, which is calculable within a variety of nuclear theory models 
\cite{Baldo1999}. Correlations between age determinations and surface temperature measurements provide an insight into 
the cooling history of isolated neutron stars, which is in turn sensitive to the EoS and the microphysics of both the crust and the core~\cite{Potekhin2015}. 

Glitches in the periodic radio signals emitted by pulsars are indicative of a rich interplay between superfluid and 
normal components in the crust of a neutron star \cite{Andersson2012,Chamel2013,PRC90Piekarewicz2014,Link2014,Haskell2015}. 
Glitching phenomena can also provide an indication of a pulsar's mass, provided that the basic microphysics of the
neutron star crust is under control  \cite{Ho2015,Pizzochero2016}. 
The densities of both the inner and outer crust are fractions of the nuclear saturation region, and one can argue 
that the nuclear energy density functional is understood to the extent that predictions in this region are under 
control \cite{NPA175BAYM1971,NPA584Pethick1995,Haensel2007,Chamel2008}. The boundary between the liquid core and the 
inhomogeneous solid crust is connected to the isospin dependence of nuclear models below saturation, as indicated by the widely used thermodynamical method \cite{PRC70Kubis2004,PRC76Kubis2007,AJ697Xu2009,PRC81Moustakidis2010,PRC85Cai2012,PRC86Moustakidis2012,PRC89Seif2014,TRRoutray}. 
A variety of different functionals (and many-body theories) have been used to 
determine the properties of the core-crust transition, including Skyrme forces \cite{AJ697Xu2009,NPA789Ducoin2007,PRC85Pearson2012,Newton2014}, 
finite-range functionals \cite{TRRoutray}, relativistic mean-field (RMF) models \cite{Horowitz:2000xj,Carriere:2002bx,PRC74Klahn2006,PRC81Moustakidis2010,Fattoyev:2010tb,PRC85Cai2012,Newton2014}, momentum-dependent interactions \cite{AJ697Xu2009,PRC86Moustakidis2012} and Brueckner--Hartree--Fock theory~\cite{Vidana2009,PRC83Ducoin2011,Li2016}. 

A key observable in the analysis of glitches is the thickness 
of the pulsar's crust \cite{Link1999,Fattoyev:2010tb,Chamel2013,PRC90Piekarewicz2014,Newton2015}, which is linked to the core-crust transition.
The crustal thickness, in turn, determines how much superfluid is available 
to pin to nuclear sites. Recent indications suggest that there is not enough superfluid in the crust for glitches to 
occur \cite{Andersson2012,Chamel2013}. 
The role of superfluid entrainment and its interplay with the lattice structure, however, is also relevant \cite{Martin2016,Watanabe2017}.
Superfluidity is key in this region and a detailed treatment based on the pairing extension of nuclear density functionals is possible \cite{grill11,pastore11,PRC85Pearson2012}. 
From a nuclear physics perspective, the description of the crust starting from a finite-range functional would allow for a description of 
the pairing channel which is free of divergences \cite{Schuck1980,Matsuo2006}. The finite-range Gogny interaction is 
constructed to reproduce the known pairing properties of nuclei \cite{Decharge1980} and is widely used in the nuclear 
structure community \cite{Robledo2011}. Its isovector properties have been analyzed by one of us in Ref.~\cite{PRC90SellahewaArnauRios2014}. 
While the symmetry energies of different Gogny forces are too soft in comparison with existing constraints \cite{Tsang2012,Lattimer2013,Lattimer2016}, a few Gogny forces do generate realistic enough equations of state. 
In the past, Gogny forces have been occasionally used in neutron-star calculations \cite{Than2011,Loan2011}. 

Here, we extend the investigation of the isovector properties of the Gogny force with an emphasis both on the higher-order contributions to the symmetry energy and on the density region relevant for the core-crust transition in neutron stars. The EoS in cold asymmetric matter with the Gogny force can be computed analytically 
\cite{PRC90SellahewaArnauRios2014}. Due to its finite-range nature and the appearance of a non-trivial exchange term, the density and asymmetry 
dependence of the EoS are expressed in terms of functions which differ from the standard polynomials that appear in the Skyrme approach and/or 
other effective descriptions \cite{NPA627Chabanat1997,NPA635Chabanat1998,NPA865XuLiChenKo2011}. In turn, this may lead to a more complex 
isospin asymmetry dependence, which we seek to identify by computing, in addition to the exact EoS, fourth- and sixth-order asymmetry effects in a Taylor expansion on the isospin asymmetry 
\cite{PRC80Chen2009,PRC85Cai2012,PRC86Moustakidis2012,PRC89Seif2014,Constantinou:2014hha,TRRoutray,Agrawal:2017sff}. The coefficients of the expansion can be computed explicitly and provide an insight on the 
importance of deviations from the standard quadratic approximation. 
In many-body calculations, for instance, the asymmetry dependence is not always directly accessible, and the parabolic approximation is often used~\cite{Vidana2009}.

We explore the accuracy of the second- and higher-order approximations for different Gogny parametrizations, by directly comparing them to the results of the exact isospin-dependent EoS.
Moreover, the use of an expansion affects the determination of the core-crust transition in neutron stars. The differences between 
the predictions extracted from the isospin expansion and from the exact EoS can be significant \cite{PRC83Ducoin2011,TRRoutray}, 
and we explore these for Gogny interactions by computing the core-crust transition
using the thermodynamical method \cite{PRC70Kubis2004,PRC76Kubis2007,AJ697Xu2009,PRC81Moustakidis2010}, which requires the thermodynamical stability of the $\beta$-equilibrated matter of  the homogeneous liquid core.
When we analyze the calculated properties of the transition point against the slope parameter $L$ of the symmetry energy in the different Gogny sets, we find an anticorrelation of the transition density with the $L$ value, whereas the transition pressure does not display a regular dependence with $L$. 

In a second stage, we compute the structure of neutron stars by solving the Tolman-Oppenheimer-Volkov (TOV)
equations using the exact EoS of the Gogny forces. We analyze the stellar mass-radius relationships for the Gogny sets that yield stable solutions of the TOV equations. We obtain the moment of inertia of the star in the slow-rotation approximation. We discuss the predictions for the dimensionless ratio $I/MR^2$ as a function of the compactness of the star, and compare with the universal fits provided by Lattimer and Schutz~\cite{Lattimer2005} and Breu and Rezzolla~\cite{Breu2016}. 
Having evaluated the transition point between the core and the crust, we can predict the thickness and mass of the crust of the neutron star. 
Among the analyzed Gogny interactions, the parameter sets D280 \cite{NPA591Blaizot1995} and, specially, D1M \cite{PRL102Goriely2009} are found to be better suited for describing the physical properties of the crust. 
Finally, we compare the predictions of these forces for the crust fraction of the moment of inertia, with the constraints deduced from observed glitches~\cite{Link1999,Andersson2012}.

The paper is structured as follows. Section~\ref{sec:NSmatter} provides a brief introduction to the properties of neutron-star matter of 
relevance for the crust, as well as to the Gogny interaction. The contributions to the symmetry energy arising from expansions on the isospin asymmetry
are studied in Sec.~\ref{sec:syme}. Following a short review on the
predictions for $\beta$-stable neutron-star matter with the Gogny force, the results for the core-crust transition are analyzed in Sec.~\ref{sec:corecrust}.
In Sec.~\ref{sec:NS}, we study the properties of neutron stars predicted by Gogny forces, with special emphasis on the crustal properties. 
We summarize our results in Sec.~\ref{sec:summ}. The appendixes contain relevant analytical formulas obtained within the 
Gogny--Hartree--Fock framework, i.e., the exact EoS and its Taylor expansion
through sixth order in the isospin asymmetry (Appendix~\ref{appendix1}), 
the chemical potentials and the pressure in isospin asymmetric matter (Appendix~\ref{appendix_p}), and the expressions for the thermodynamical potential used to locate the core-crust transition (Appendix~\ref{appendix_thermal}).

\section{Formalism}
\label{sec:NSmatter}

The Gogny two-body effective nuclear interaction  \cite{Decharge1980} used in the present work is given 
(neglecting the spin-orbit force, which vanishes in nuclear matter) by
\begin{eqnarray}\label{VGogny}
  V (\mathbf{r}_1 , \mathbf{r}_2) &=&  \sum_{i=1,2} \left( W_i + B_i P_\sigma - H_i P_\tau - M_i P_\sigma P_\tau \right)e^{-r^2 /\mu_i^2}  
\nonumber
   \\
 && \mbox{} + t_3 \left( 1+x_3 P_\sigma \right) \rho^\alpha(\mathbf{R})\delta(\mathbf{r}) .
\end{eqnarray}
The two-body spin-exchange and isospin-exchange operators are denoted by $P_\sigma$ and 
$P_\tau$, respectively; $\mathbf{r} =\mathbf{r}_1 - \mathbf{r}_2 $ is the relative distance between two nucleons; and 
$\mathbf{R}= (\mathbf{r}_1 + \mathbf{r}_2)/2$
 is the center of mass coordinate. 
 The first term in Eq.~(\ref{VGogny}) is modulated by two Gaussians, with short- and long-range parameters, 
 $\mu_i$.  The second term is a zero-range density dependent contribution. 
The coefficients $t_3$, $x_3$, $W_i$,
 $B_i$, $H_i$ and $M_i$ ($i=1, 2$) are the fit parameters of the interaction (in principle, the ranges $\mu_i$ and the $\alpha$ power are also parameters, but in practice they are fixed in the fitting procedure).

For a given nuclear interaction, the energy per baryon $E_b(\rho, \delta)$ in asymmetric infinite nuclear matter 
can be written as a function of the total baryon number density $\rho = \rho_n + \rho_p$ and 
of the isospin asymmetry $\delta= (\rho_n - \rho_p)/\rho$, 
where $\rho_n$ and $\rho_p$ are, respectively, the neutron and proton number densities.
The analytical expression of $E_b (\rho, \delta)$ for the Gogny interaction is provided in Eqs.~(\ref{eq:eb.terms})--(\ref{eq:eb.exch}) of Appendix \ref{appendix1}. 
It is also common to express the energy per baryon as a Taylor expansion with respect to the isospin asymmetry around $\delta=0$:
\begin{eqnarray}\label{EoS}
 E_b(\rho, \delta) &=& E_b(\rho, \delta=0) + E_{\mathrm{sym}, 2}(\rho) \delta^2 + E_{\mathrm{sym}, 4 }(\rho)\delta^4 + ... \nonumber
 \\
 && \mbox{} + E_{\mathrm{sym}, 2k}(\rho)\delta^{2k} + \mathcal{O}(\delta^{2k+2}) \, .
\end{eqnarray}
Charge symmetry of the nuclear forces is assumed, so that the strong interaction is symmetric under neutron and proton exchange and only 
even powers of $\delta$ appear in Eq.~(\ref{EoS}). 
The first coefficient in this expansion, $E_b(\rho, \delta=0)$, gives the energy per baryon 
in symmetric nuclear matter. The symmetry energy coefficient is usually defined as the second-order coefficient in the expansion, 
$E_{\mathrm{sym}, 2} (\rho)$. Another popular notation for $E_{\mathrm{sym}, 2} (\rho)$ in the literature is $S(\rho)$
 \cite{PRC90SellahewaArnauRios2014,Piekarewicz:2008nh}.
If the isospin dependence of the EoS is rich, however, one expects that the higher-order coefficients may provide relatively important corrections \cite{AJ697Xu2009,PRC80Chen2009}. 
In general, the symmetry energy coefficients at a given order $2k$ in the isospin asymmetry are defined as
\begin{equation}\label{esym}
 \left. E_{\mathrm{sym}, 2k} (\rho) = \frac{1}{(2k)!} \frac{\partial^{2k} E_b(\rho, \delta)}{\partial \delta^{2k}}\right|_{\delta=0} \, . 
\end{equation}
These coefficients are intimately related to the isospin dependence of the nuclear interaction, and are directly connected to the properties 
of the single-nucleon potential in asymmetric systems \cite{NPA865XuLiChenKo2011,PRC85Chen2012}. 
Analytical expressions for the symmetry energy coefficients $E_{\mathrm{sym},2}(\rho)$, $E_{\mathrm{sym},4}(\rho)$, and $E_{\mathrm{sym},6}(\rho)$ for the Gogny interaction
are given in Eqs.~(\ref{eq:esym2})--(\ref{eq:esym6}) of Appendix~\ref{appendix1}.
Recent calculations in many-body perturbation theory have shown  
that the isospin asymmetry expansion~(\ref{EoS}) may not be convergent at zero temperature 
when the many-body corrections beyond the Hartree-Fock mean-field level are incorporated \cite{Wellenhofer2016}. 
We do not deal with this complication here since we will be working at the Hartree-Fock level, 
where no non-analyticities are found in the equation of state.

In our applications of Gogny forces to calculations of the matter of the core of neutron stars,
we shall consider neutron star cores consisting of $\beta$-stable $\textit{npe}$ asymmetric nuclear matter. This is the expected composition of the neutron star core below the inner edge of the crust. The very dense inner core of the star may harbor more exotic particles such as hyperons \cite{Fortin:2014mya,Tolos:2016hhl}.
However, in this paper we are mainly interested in studying the properties of the nucleonic EoS of Gogny forces and, hence, we shall avoid dwelling with cores with more exotic components. In the system of $\textit{npe}$ matter, we can express the total energy density as the sum of the baryon and electron contributions, i.e., 
\begin{equation}
 \mathcal{H} (\rho, \delta)= \mathcal{H}_b (\rho, \delta)+ \mathcal{H}_e (\rho, \delta).
\end{equation}
The baryon contribution includes the energy per baryon $E_b (\rho, \delta)$ as well as the nucleon rest mass $m$:
\begin{equation}
 \mathcal{H}_b (\rho, \delta) = \rho E_b (\rho, \delta) + \rho m \, .
\end{equation}
It is to be noted that we use natural units $\hbar=c=1$.
The electronic contribution is that of a relativistic degenerate free Fermi gas \cite{AstrophysicalJ170BaymPethick1971}:
\begin{align}
 \mathcal{H}_e = \frac{m_e^4}{8\pi^2} & \left[ x_F \sqrt{1+x_F^2} \left(2 x_F^2 + 1 \right) - \arcsinh (x_F) \right],
\end{align}
where $m_e$ is the mass of the electron and the dimensionless Fermi momentum is $x_F \equiv k_{Fe}/m_e = (3 \pi^2 \rho_e)^{1/3}/m_e$, 
with $\rho_e$ being the electron number density. We impose charge neutrality and thus consider $\rho_e=\rho_p$.

The pressure of the system contains the baryon and electron contributions,
\begin{equation}\label{eq:pressure}
 P (\rho, \delta) = P_b (\rho, \delta) + P_e (\rho, \delta) \, ,
\end{equation}
with
\begin{equation}\label{eq:Pb}
 P_b= \rho^2 \frac{\partial E_b}{\partial \rho} \hspace{1cm}\mathrm{and} \hspace{1cm} P_e= \rho_e^2 \frac{\partial E_e}{\partial \rho_e},
\end{equation}
where $E_e$ is the electron energy per particle. 
An analytical expression for the baryon pressure in Gogny interactions is provided in Appendix~\ref{appendix_p}. The corresponding electron pressure is
\begin{align}
 P_e &= \frac{m_e^4}{24\pi^2} \left[ x_F \sqrt{1+x_F^2} \left(2 x_F^2 - 3 \right) + \arcsinh (x_F) \right].
 \label{eq:press_el}
\end{align}
We denote by $\mu_n$, $\mu_p$, and $\mu_e$ the chemical potentials of neutrons, protons, and electrons, respectively. 
The electron chemical potential is
\begin{equation}\label{eq:potentialele}
 \mu_e = \frac{\partial \mathcal{H}_e}{\partial \rho_e} = \sqrt{k_{Fe}^2 + m_e^2} = \sqrt{ (3 \pi^2 \rho_e)^{2/3} + m_e^2 } \, .
\end{equation}
For neutrons and protons, chemical potentials are obtained from density derivatives of the energy density,
\begin{equation} \label{eq:potentials}
 \mu_n = \frac{\partial \mathcal{H}_b}{\partial \rho_n} \hspace{1cm} \mathrm{and} \hspace{1cm} \mu_p = \frac{\partial \mathcal{H}_b}{\partial \rho_p} \, ,
\end{equation}
or, alternatively, from the single-particle potentials at the corresponding Fermi surfaces \cite{PRC90SellahewaArnauRios2014}. 
Analytical expressions for the nucleon chemical potentials are given in Appendix~\ref{appendix_p}.
With Eqs.~(\ref{eq:potentialele}) and~(\ref{eq:potentials}), we can write the 
pressures in Eq.~(\ref{eq:Pb}) as 
\begin{eqnarray}
\label{PbPe}
 P_b (\rho, \delta) &=& \mu_n\rho_n + \mu_p\rho_p - \mathcal{H}_b (\rho, \delta) ,
\nonumber \\[2mm]
 P_e (\rho, \delta) &=& \mu_e \rho_e - \mathcal{H}_e (\rho, \delta).
\end{eqnarray}

Before we proceed to study the core-crust transition in neutron stars with the Gogny interaction, 
in the following Sec.~\ref{sec:syme} we analyze the symmetry energy of Gogny forces at higher orders. 
If the exact EoS is replaced by its Taylor expansion at second and higher orders in the isospin asymmetry, 
the properties of the core-crust transition may be affected. Hence, in Sec.~\ref{sec:corecrust} when we study 
the neutron star core-crust transition we will also analyze the errors introduced by breaking the isospin 
asymmetry expansion of the EoS at finite orders. Finally, in Sec.~\ref{sec:NS} we will compute with the 
Gogny forces global properties, such as masses and sizes, of neutron stars and their crusts.

\section{Symmetry energy of Gogny forces}
\label{sec:syme}

There are about ten available Gogny parametrizations in the literature \cite{PRC90SellahewaArnauRios2014}. 
In our calculations, we use the interactions D1 \cite{Decharge1980}, D1S \cite{Berger1991},
 D1M \cite{PRL102Goriely2009}, D1N \cite{PLB668Chappert2008}, and the family of forces 
 D250, D260, D280, and D300 \cite{NPA591Blaizot1995}. 
D1 is the original Gogny force and was fit to the properties of a few closed-shell nuclei and of nuclear matter at saturation \cite{Decharge1980}. 
D1S was introduced some years later with a focus on describing nuclear fission \cite{Berger1991}, and remains the most widely used Gogny force to date.
The models D250, D260, D280, and D300 were devised to have different nuclear matter compression moduli for calculations of the breathing mode in nuclei \cite{NPA591Blaizot1995}. 
D1N is a revised parametrization of D1S that aims to improve on some of its features, such as the isotopic trends of binding energies \cite{PLB668Chappert2008}. 
The isospin dependence of this force was calibrated by considering the Friedman-Pandharipande neutron matter EoS in the  density region from subsaturation up to saturation densities \cite{PLB668Chappert2008}. 
Finally, D1M \cite{PRL102Goriely2009} has been conceived as a high-accuracy nuclear mass model within the Hartree--Fock--Bogoliubov approach. 
The D1M parameters have been obtained by a global fit to essentially all measured nuclear masses, while keeping the properties of nuclear matter and neutron matter in satisfactory agreement with realistic many-body calculations of the EoS \cite{PRL102Goriely2009}. 
We note that, as discussed in Ref.~\cite{PRC90SellahewaArnauRios2014}, none of these Gogny parametrizations 
fall within the low-density ($\rho < 0.08$ fm$^{-3}$) microscopic predictions based on chiral effective field theory
proposed in Ref.~\cite{Brown2014}. At densities between $0.10$ fm$^{-3}$ and $0.17$ fm$^{-3}$, 
however, Gogny parametrizations overlap with the microscopic constraints of Ref.~\cite{Brown2014}.

\subsection{Second-, fourth-, and sixth-order contributions to the symmetry energy}
\begin{figure}[t!]
 \centering
 \includegraphics[width=1\linewidth, clip=true]{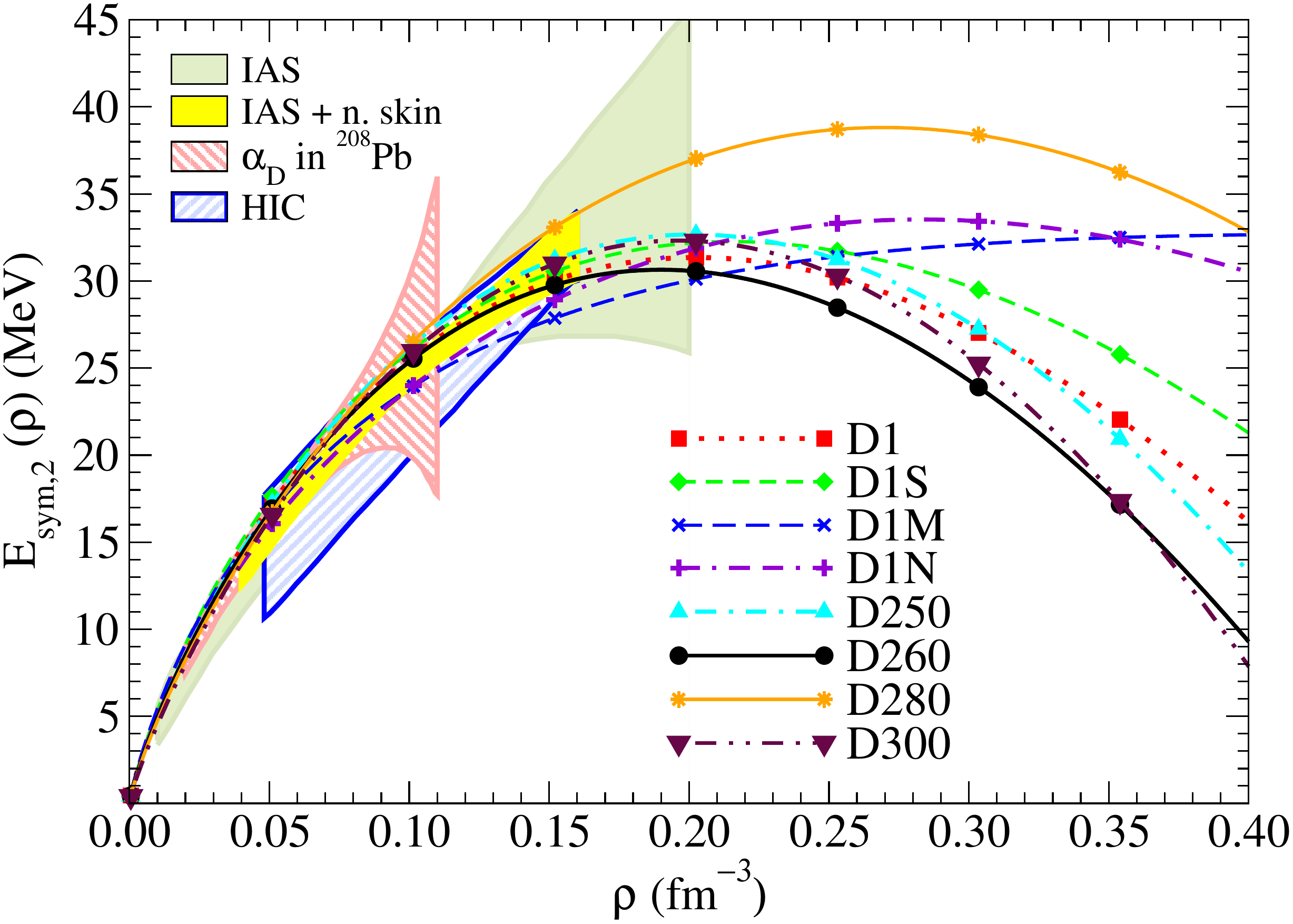}
 \caption{Density dependence of the second-order symmetry energy coefficient $E_{\mathrm{sym}, 2}(\rho)$ for different Gogny interactions. 
Also represented are the symmetry energy constraints extracted from the analysis of data on isobaric analog states (IAS)
and of IAS data combined with neutron skins (IAS+n.skin) \cite{Danielewicz:2013upa}, 
the constraints from the electric dipole polarizability in lead ($\alpha_D$ in $^{208}$Pb) \cite{Zhang:2015ava},
and from transport simulations of heavy-ion collisions of tin isotopes (HIC) \cite{Tsang:2008fd}.}
\label{fig:esym2}
\end{figure}

We first analyze the second- and higher-order terms (symmetry energy coefficients) in the Taylor expansion, Eq.~(\ref{EoS}), of the energy per particle for the Gogny interaction. A detailed characterisation of these terms is useful in order to understand the calculations of $\beta$-equilibrium matter as well as the core-crust transition. 
In Fig.~\ref{fig:esym2} we show the second-order symmetry energy coefficient $E_{\mathrm{sym}, 2} (\rho)$ for all 
the considered Gogny interactions. At low densities $\rho \lesssim 0.1$ fm$^{-3}$, $E_{\mathrm{sym}, 2} (\rho)$ has comparable values in all the forces and 
increases with density. From $\rho \gtrsim 0.1$ fm$^{-3}$ on, there are substantial differences between the predictions of different 
parametrizations. 
In comparison with existing empirical constraints for the symmetry energy at subsaturation density \cite{Danielewicz:2013upa,Zhang:2015ava,Tsang:2008fd}, 
one finds that the Gogny functionals in general respect them (cf.\ Fig.~\ref{fig:esym2}).
At saturation density, $E_{\mathrm{sym}, 2} (\rho)$ of the Gogny forces lies between 28.5 and 33 MeV.
The relatively flat density dependence of $E_{\mathrm{sym}, 2} (\rho)$ around saturation for all the interactions in turn translates into a relatively small slope parameter $L$, as discussed in Ref.~\cite{PRC90SellahewaArnauRios2014} (see also Table~\ref{Table-saturation} below). As a general trend, all curves peak at values $E_{\mathrm{sym}, 2} (\rho) \sim 30-40$ MeV right above saturation density, with a subsequent flattening.
Beyond this maximum, all parametrizations yield a symmetry energy that decreases with density
(in D1M, though, this happens only at substantially high densities).
In all cases, $E_{\mathrm{sym}, 2} (\rho)$ beyond $0.4$ fm$^{-3}$ eventually becomes negative (in D1M only at a very large density of $1.9$ fm$^{-3}$), signaling the onset of an isospin instability. We do not consider explicitly the effect of this instability in the following discussions.

\begin{figure}[t!]
 \centering
 \includegraphics[width=1\linewidth, clip=true]{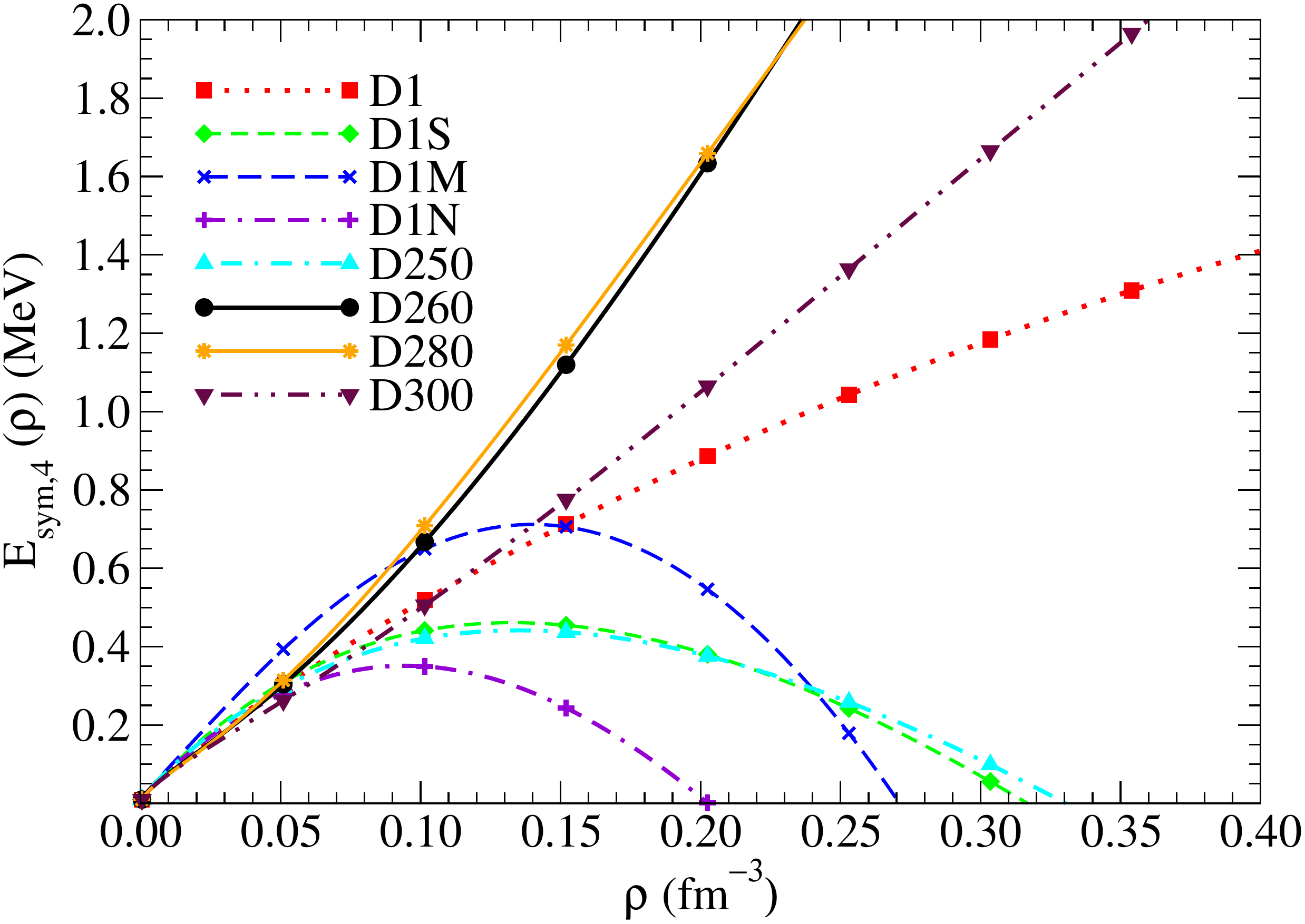}
 \caption{Density dependence of the fourth-order symmetry energy coefficient $E_{\mathrm{sym}, 4}(\rho)$ 
 for different Gogny interactions.}
\label{fig:esym4}
\end{figure}

\begin{figure}[t!]
 \centering
 \includegraphics[width=1\linewidth, clip=true]{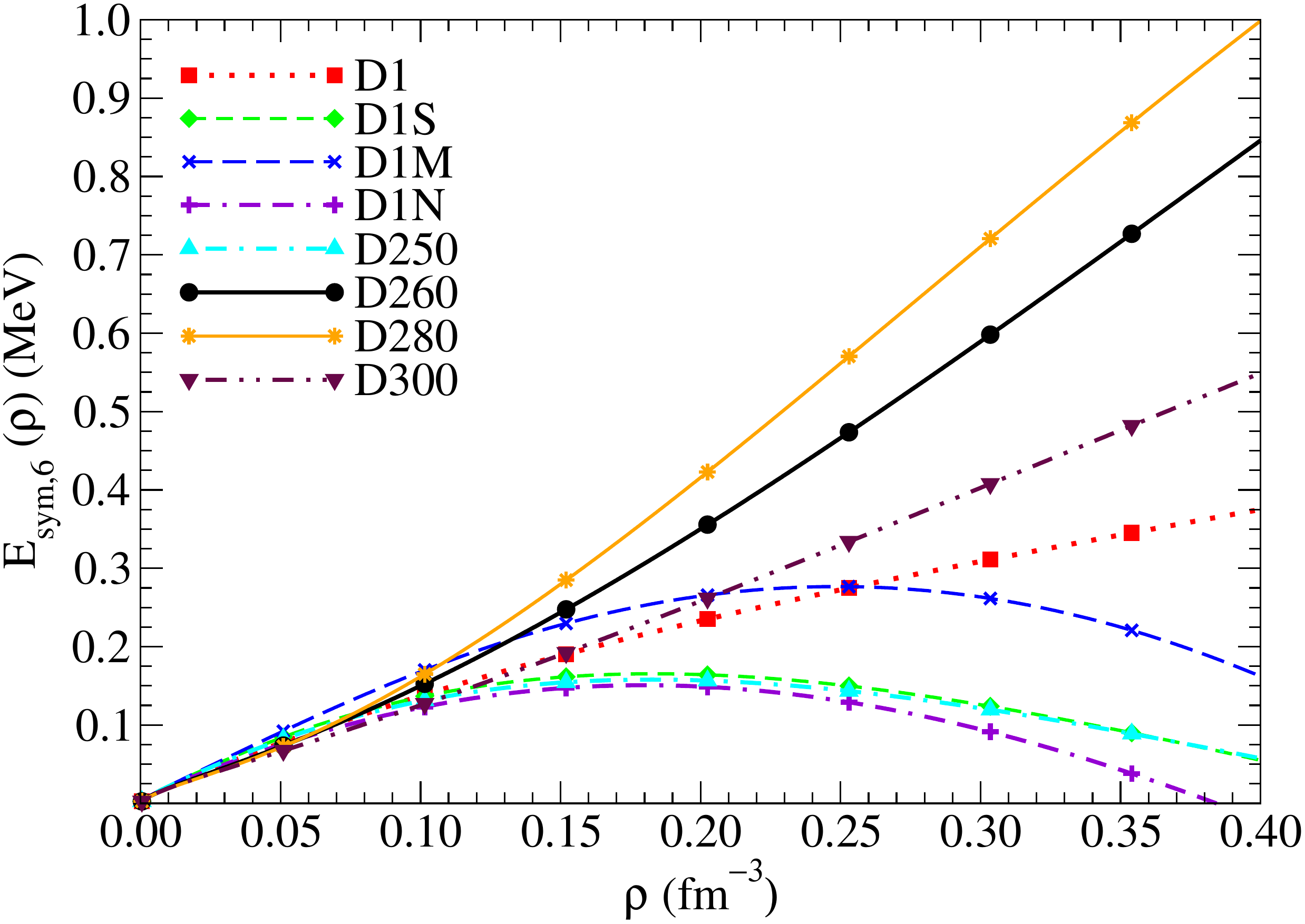}
 \caption{Density dependence of the sixth-order symmetry energy coefficient $E_{\mathrm{sym}, 6}(\rho)$
 for different Gogny interactions.}
\label{fig:esym6}
\end{figure}

We show the symmetry energy coefficients of fourth-order, $E_{\mathrm{sym}, 4}(\rho)$, and sixth-order, $E_{\mathrm{sym}, 6}(\rho)$, in Figs.~\ref{fig:esym4} and \ref{fig:esym6}, respectively. 
On the one hand, at subsaturation densities both terms are relatively small: below saturation density, $E_{\mathrm{sym}, 4}(\rho)$ is below $\approx$\,1 MeV and 
$E_{\mathrm{sym}, 6}(\rho)$ does not go above $\approx$\,0.3 MeV. These values can be compared with the larger values of $E_{\mathrm{sym}, 2}(\rho) > 10$ MeV in the same density regime. 
One should also consider that in the expansion of Eq.~(\ref{EoS}) the terms $E_{\mathrm{sym}, 4}(\rho)$ and $E_{\mathrm{sym}, 6}(\rho)$ carry additional factors $\delta^2$ and $\delta^4$ with respect to $E_{\mathrm{sym}, 2}(\rho)$, and their overall magnitude will therefore be smaller.
On the other hand, above saturation density, we observe two markedly different behaviors for the density dependence of $E_{\mathrm{sym}, 4}$ and $E_{\mathrm{sym}, 6}$. 
For both $E_{\mathrm{sym}, 4}(\rho)$ and $E_{\mathrm{sym}, 6}(\rho)$, we find a group of parametrizations (D1S, D1M, D1N, and D250) that reach a maximum and then decrease with density. 
We call this set of forces {\sl ``group~1"} from now on. A second set of forces, {\sl ``group~2''}, is formed of D1, D260, D280, and D300,
which yield $E_{\mathrm{sym}, 4}(\rho)$ and $E_{\mathrm{sym}, 6}(\rho)$ terms
that do not reach a maximum and increase steeply in the range of the studied densities.

The difference in density dependence between the second-order symmetry energy and its higher-order corrections can be understood by 
decomposing them into terms associated to the different contributions from the nuclear Hamiltonian.
All three coefficients $E_{\mathrm{sym}, 2}$, $E_{\mathrm{sym}, 4}$ and $E_{\mathrm{sym}, 6}$ include a 
kinetic component, which decreases substantially as the order increases. The
$E_{\mathrm{sym}, 2}$ coefficient also receives contributions 
from the zero-range term of the force [Eq.~(\ref{eq:eb.zr})] as well as from the finite-range direct and exchange terms
[Eqs.~(\ref{eq:eb.dir}) and (\ref{eq:eb.exch})]:
\begin{align}
E_{\mathrm{sym}, 2} (\rho) &= \frac{\hbar^2}{6m} \left(  \frac{3 \pi^2}{2}\right)^{2/3} \rho^{2/3} \nonumber \\
& - \frac{1}{8} t_3 \rho^{\alpha+1} (2x_3 +1) + \frac{1}{2} \sum_{i=1,2} \mu_i^3 \pi^{3/2}  {\cal B}_i  \rho \nonumber
\\
& +\frac{1}{6}\sum_{i=1,2}  \left[-{\cal C}_i  G_1 (\mu_i k_F)+ {\cal D}_i G_2 (\mu_i k_F)  \right] \, .
\end{align}
The expressions for the constants ${\cal B}_i$, ${\cal C}_i$, and ${\cal D}_i$ and the $G_n (\mu_i k_F)$ functions are given in Appendix~\ref{appendix1}.
We note that the direct terms of the finite-range contribution to $E_{\mathrm{sym}, 2}$ are directly proportional to the constants ${\cal B}_i$ and to the density $\rho$. 
The functions $G_n (\mu_i k_F)$ are due solely to the exchange contribution in the matrix elements of the Gogny force. 
One can equally say that they reflect the contribution of the momentum dependence of the interaction to the symmetry energy. 
As discussed in Ref.~\cite{PRC90SellahewaArnauRios2014}, the zero-range term, the direct term, and the exchange (momentum-dependent) term contribute 
with similar magnitudes to the determination of $E_{\mathrm{sym}, 2}$ with Gogny forces. However, they contribute with different signs, which leads to 
cancellations in $E_{\mathrm{sym}, 2}$ between the power-law zero-range term, the linear density-dependent direct term, and the more complex exchange term. 
Depending on the parametrization, the sum of the zero-range and direct terms is positive and the exchange term is negative, 
or the other way around. In any case, there is a balance between terms, 
which gives rise to a somewhat similar density dependence of the symmetry energy coefficient $E_{\mathrm{sym}, 2}$ for all parameter sets. 

In contrast to the case of the $E_{\mathrm{sym}, 2}$ coefficient, neither the zero-range nor the direct term contribute to 
the $E_{\mathrm{sym}, 4}$ and $E_{\mathrm{sym}, 6}$ coefficients, 
\begin{align}
E_{\mathrm{sym}, 4} (\rho) &= \frac{\hbar^2}{162m} \left(  \frac{3 \pi^2}{2}\right)^{2/3} \rho^{2/3} \nonumber \\
& +\frac{1}{324} \sum_{i=1,2} \left[ {\cal C}_i  G_3 (\mu_i k_F)  + {\cal D}_i G_4 (\mu_i k_F) \right] \, ,
\nonumber \\
E_{\mathrm{sym}, 6} (\rho)  &= \frac{7\hbar^2}{4374m} \left(  \frac{3 \pi^2}{2}\right)^{2/3} \rho^{2/3} \nonumber \\
& + \frac{1}{43740} \sum_{i=1,2} \left[ {\cal C}_i  G_5(\mu_i k_F)  - {\cal D}_i  G_6(\mu_i k_F) \right] \, ,
\label{eq:esym46}
\end{align}
because both the zero-range and the direct components of the energy per particle [cf.\ Eqs.~(\ref{eq:eb.zr}) and (\ref{eq:eb.dir})] depend on the square of the isospin asymmetry, $\delta^2$. 
In other words, the higher-order corrections to the symmetry energy are only sensitive to the kinetic term and to 
the momentum-dependent term, i.e., the exchange term of the Gogny force. 
We note that the same pattern is found in zero-range Skyrme forces. That is, also in Skyrme forces the higher-order symmetry energy coefficients
$E_{\mathrm{sym}, 4}$, $E_{\mathrm{sym}, 6}$, etc., arise exclusively from the kinetic term and from the momentum-dependent term 
of the interaction, which in the Skyrme forces is the term with the usual $t_1$ and $t_2$ parameters \cite{NPA627Chabanat1997,NPA635Chabanat1998}.
In the Skyrme case, though, the functional dependence of the momentum-dependent contribution to the symmetry energy coefficients is   
proportional to $\rho^{5/3}$, whereas in the Gogny case it has a more intricate density dependence due to the finite range of the interaction,
which is reflected in the $G_n (\mu_i k_F)$ functions.

In both $E_{\mathrm{sym}, 4}(\rho)$ and $E_{\mathrm{sym}, 6}(\rho)$ of Gogny forces, cf.\ Eq.~(\ref{eq:esym46}), the exchange term is given by the product of two parametrization-dependent constants, ${\cal C}_i$ and ${\cal D}_i$, and two density-dependent functions, 
$G_3$ and $G_4$, or $G_5$ and $G_6$. Because the density dependence of these functions is similar, one does expect 
that comparable density dependencies arise for the fourth- and the sixth-order, as observed in Figs.~\ref{fig:esym4} and \ref{fig:esym6}.
This simple structure also provides an explanation for the appearance of two distinct groups of forces in terms of the density dependence 
 of $E_{\mathrm{sym}, 4}(\rho)$ and $E_{\mathrm{sym}, 6}(\rho)$. In ``group~1'' forces, the fourth- and sixth-order contributions to the symmetry 
energy change signs as a function of density, whereas ``group~2'' forces produce monotonically increasing functions of density. The change of sign
 is necessarily due to the exchange contribution, which in the case of ``group~1'' forces must also be attractive 
 enough to overcome the kinetic term. 

\begin{figure}[t!]
 \centering
 \includegraphics[width=1\linewidth, clip=true]{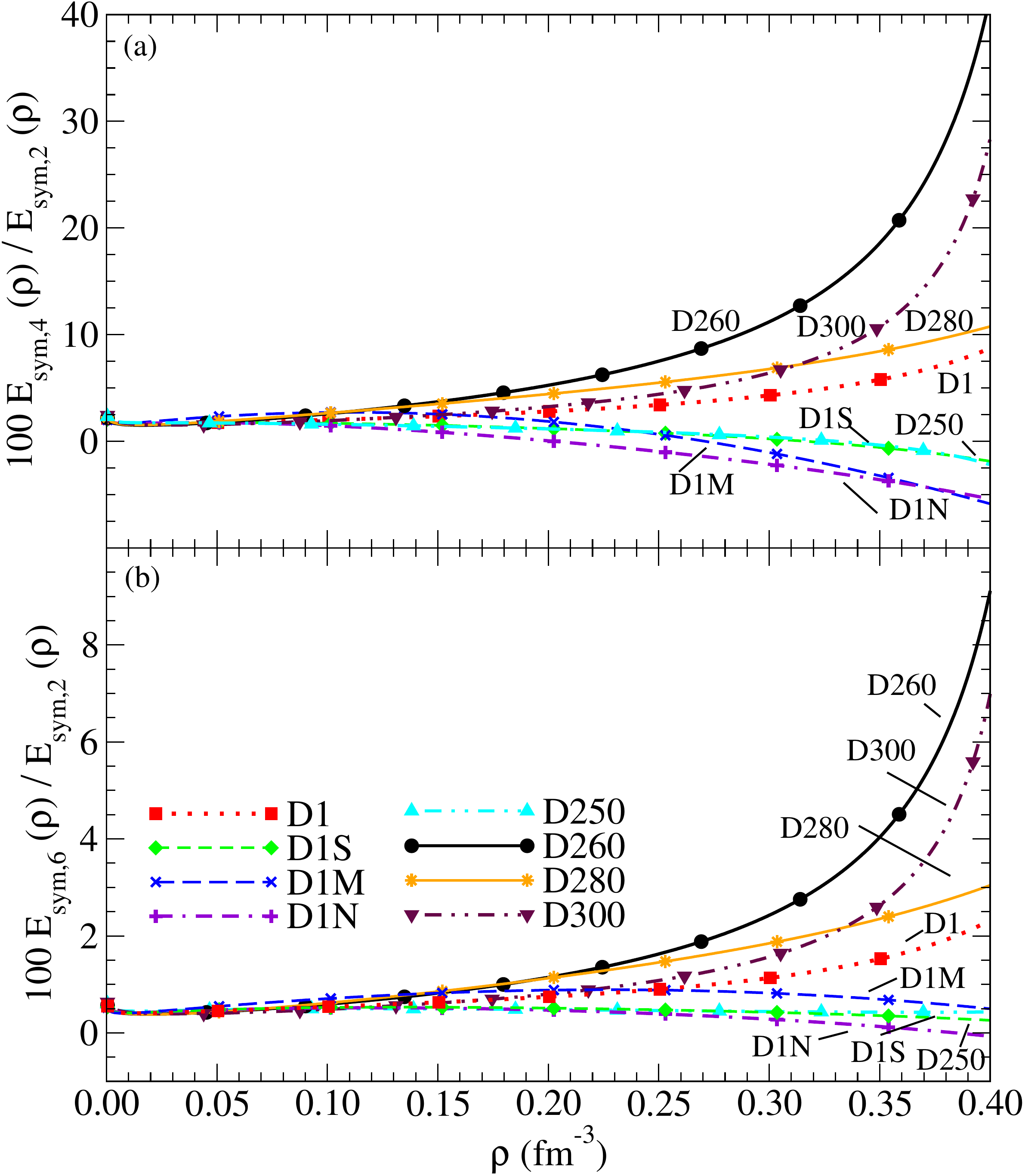}
  \caption{Density dependence of the ratios $E_{\mathrm{sym}, 4} (\rho)/E_{\mathrm{sym}, 2} (\rho)$ (top panel) 
  and $E_{\mathrm{sym}, 6} (\rho)/E_{\mathrm{sym}, 2} (\rho)$ (bottom panel) for different Gogny interactions.}
\label{fig:E46E2_Gogny}
\end{figure}

\begin{table*}[t!]
\begin{ruledtabular}
\begin{tabular}{ddddddddd}	
 \multicolumn{1}{c}{Force}  &  \multicolumn{1}{c}{D1}  &  \multicolumn{1}{c}{D1S}  &  \multicolumn{1}{c}{D1M}  &  \multicolumn{1}{c}{D1N}  &  \multicolumn{1}{c}{D250}  &  \multicolumn{1}{c}{D260}  &  \multicolumn{1}{c}{D280}  &  \multicolumn{1}{c}{D300}  \\
 \hline
 \multicolumn{1}{c}{$\rho_0$}  &   0.167    &   0.163    &   0.165    &   0.161    &   0.158    &   0.160    &   0.153    &   0.156    \\
 \multicolumn{1}{c}{$E_0$}  &   -16.31   &   -16.01   &   -16.03   &   -15.96   &   -15.80   &   -16.26   &   -16.33   &   -16.22   \\
 \multicolumn{1}{c}{$K_0$}  &   229.37   &   202.88   &   224.98   &   225.65   &   249.41   &   259.49   &   285.20   &   299.14   \\ \hline
 \multicolumn{1}{c}{$E_{\mathrm{sym}, 2}$($\rho_0$)}  &   30.70    &   31.13    &   28.55    &   29.60    &   31.54    &   30.11    &   33.14    &   31.23    \\
 \multicolumn{1}{c}{$E_{\mathrm{sym}, 4}$($\rho_0$)}  &   0.76     &   0.45     &   0.69     &   0.21     &   0.43     &   1.20     &   1.18     &   0.80     \\
 \multicolumn{1}{c}{$E_{\mathrm{sym}, 6}$($\rho_0$)}  &   0.20     &   0.16     &   0.24     &   0.15     &   0.16     &   0.27     &   0.29     &   0.20     \\
 \multicolumn{1}{c}{$L$}  &   18.36    &   22.43    &   24.83    &   33.58    &   24.90    &   17.57    &   46.53    &   25.84    \\
 \multicolumn{1}{c}{$L_4$}  &   1.75     &   -0.52    &   -1.04    &   -1.96    &   -0.33    &   4.73     &   4.36     &   2.62     \\
 \multicolumn{1}{c}{$L_6$}  &   0.46     &   0.08     &   0.42     &   0.08     &   0.09    &   0.99     &   1.19     &   0.63     \\
 \hline
 \multicolumn{1}{c}{$E_\mathrm{sym}^{PA}$($\rho_0$)}  &   31.91    &   31.95    &   29.73    &   30.14    &   32.34    &   31.85    &   35.89    &   32.44    \\
 \multicolumn{1}{c}{$L_{PA}$}  &   21.16    &   22.28    &   24.67    &   31.95    &   24.94    &   24.33    &   53.25    &   29.80    \\
\end{tabular}
\end{ruledtabular}
\caption{Saturation properties of nuclear matter studied using Gogny interactions. The saturation density $\rho_0$ has units of fm$^{-3}$ and all other properties have units of MeV.}
\label{Table-saturation}
\end{table*}

For further insight into the relevance of $E_{\mathrm{sym}, 4} (\rho)$ and $E_{\mathrm{sym}, 6} (\rho)$ for the Taylor expansion of the EoS at each density, we plot in Fig.~\ref{fig:E46E2_Gogny} their ratios with respect to $E_{\mathrm{sym}, 2} (\rho)$. In the zero density limit, we see that both ratios tend to a constant value. This is expected in the non-interacting case, although the actual values of these ratios are modified by the exchange contributions. In this limit, we find $E_{\mathrm{sym}, 4}/E_{\mathrm{sym}, 2} \approx 1.5 \%$ and $E_{\mathrm{sym}, 6}/E_{\mathrm{sym}, 2} \approx 0.4 \%$. 
At low, but finite densities, $\rho \lesssim 0.1$ fm$^{-3}$, the ratio $E_{\mathrm{sym}, 4}/E_{\mathrm{sym}, 2}$ is relatively flat and not larger than $3\%$. The ratio for the sixth-order term is also mildly density-dependent and less than $0.6 \%$. Beyond saturation, both ratios increase in absolute value, to the point that for some parametrizations the ratio of the fourth- (sixth-) order term to the second-order term is not negligible and of about $10-30 \%$ ($2-8 \%$) or even more. In particular, this is due to the decreasing trend of $E_{\mathrm{sym}, 2} (\rho)$ with increasing density for several interactions when $\rho$ is above saturation.
We may compare these results for the ratios with previous literature. For example, in the calculations of Ref.~\cite{PRC86Moustakidis2012} with the momentum-dependent interaction (MDI) and with the Skyrme forces SLy4, SkI4 and Ska, we find values $ \left| E_{\mathrm{sym}, 4} (\rho)/E_{\mathrm{sym}, 2} (\rho) \right| < 8 \%$ at $\rho \sim 0.4 $ fm$^{-3}$. In the same reference, we find that the Thomas-Fermi model of Myers and Swiatecki yields a ratio $ \left| E_{\mathrm{sym}, 4} (\rho)/E_{\mathrm{sym}, 2} (\rho) \right|$ reaching 60\% already at moderate density $\rho= 1.6\rho_0$ \cite{PRC86Moustakidis2012}.
With RMF models such as FSUGold or IU-FSU, at densities $\rho \sim 0.4 $ fm$^{-3}$ one has ratios $ \left| E_{\mathrm{sym}, 4} (\rho)/E_{\mathrm{sym}, 2} (\rho) \right| < 4 \%$ \cite{PRC85Cai2012}. All in all, it appears that Gogny parametrizations provide ratios that are commensurate with previous literature. 

\subsection{Isovector properties at saturation}

The physics of the core-crust transition occurs at sub-saturation densities, which is also the finite nucleus 
regime where Gogny forces are fit to. At slightly higher densities, at and around saturation, one also expects 
the isovector properties to be relatively well under control \cite{Vidana2009}. Large deviations between functionals
at saturation would point to large systematic uncertainties in the nuclear density functional \cite{Dobaczewski2014}. 

We present the isoscalar saturation properties for Gogny functionals in the first three rows of Table \ref{Table-saturation}. 
The saturation density is close to $\rho_0 \simeq 0.16$ fm$^{-3}$ in all cases. The 
saturation energy is also within a few percent of the standard value $E_0 \simeq -16$ MeV. The 
compressibility $K_0$ describes the curvature of the energy per particle
around the saturation point, and has a wider range of values, $202 \lesssim K_0 \lesssim 300$ MeV. Part of this 
variation is due to the family of interactions D250--D300, which were specifically designed to have a range 
of nuclear compressibilities \cite{NPA591Blaizot1995}. The lower bound, however, is given by the D1S force 
\cite{Berger1991}, with $K_0 \approx 202$ MeV. On the whole, isoscalar saturation properties are in line with expectations. 

Rows 4 to 6 of Table \ref{Table-saturation} include the symmetry energy coefficients at 
the saturation density calculated at second, fourth and sixth order. The second-order 
symmetry energy coefficient is in the range of $E_{\mathrm{sym},2} (\rho_0) \approx 
28-33$~MeV, which agrees well with known empirical and theoretical values 
\cite{Tsang2012,Lattimer2013,Lattimer2016}. This range may also be compared with the values 
derived from recent microscopic calculations, such as e.g.\ the range of 28.5--33.3 MeV 
proposed from ab initio calculations of the electric dipole polarizability in $^{48}$Ca \cite{Birkhan:2016qkr} 
using chiral interactions \cite{PRC83Hebeler2011,PRC91Ekstrom2015}, 
and the ranges 28--35 MeV \cite{PRC95Holt2017} and 29--34 MeV \cite{Drischler1710.08220} 
from nuclear and neutron matter calculations from chiral effective field theory. For the 
symmetry energy corrections of fourth and sixth order in the Gogny forces, we find 
values of $E_{\mathrm{sym}, 4} (\rho_0) \approx 0.2-1.2$ MeV and $E_{\mathrm{sym}, 6} 
(\rho_0) \approx 0.15-0.3$ MeV at saturation density. 
Clearly, they exhibit larger relative variations in the different forces than 
$E_{\mathrm{sym},2} (\rho_0)$.
Bulk isovector properties are hardly ever considered in the fit procedure of Gogny 
interactions. The isospin dependence of these forces is guided by fits to finite-nucleus 
properties, close to isospin-symmetric conditions. It is therefore not surprising to 
find large variations in the isovector properties predicted by different 
parametrizations, in contrast to the well-constrained isoscalar properties.

If we expand $E_{\mathrm{sym}, 2k} (\rho)$ around saturation density $\rho_0$, we obtain the expression
\begin{equation}
 E_{\mathrm{sym}, 2k} (\rho)= E_{\mathrm{sym}, 2k} (\rho_0) + L_{2k} \epsilon + \mathcal{O}(\epsilon^2),
 \label{eq:e_l2k}
\end{equation}
where $ \epsilon = (\rho - \rho_ 0)/(3\rho_0)$ is the relative density variation with respect to $\rho_0$. 
The slope parameters $L_{2k}$ are given by
\begin{equation}\label{eq:L2k}
 \left. L_{2k}= 3 \rho_0 \frac{\partial E_{\mathrm{sym}, 2k} (\rho)}{\partial \rho} \right|_{\rho_0}.
\end{equation}
Recalling Eq.~(\ref{EoS}) and the saturation condition of nuclear forces, we see that the density slope at saturation of the energy per particle $E_b (\rho, \delta)$ of asymmetric matter can be parametrized as
\begin{equation}
\left. \frac{\partial E_b (\rho, \delta)}{\partial \rho} \right|_{\rho_0} =
\frac{1}{3\rho_0} \left( L_2 \delta^2 + L_4 \delta^4 + L_6 \delta^6 + \cdots \right) .
 \label{eq:slope2k}
\end{equation}
$L_2$ is usually referred to as the slope parameter of the symmetry energy and is denoted as $L$, which we do from here onwards. 

The expressions for $L$, $L_{4}$, and $L_{6}$ in the Gogny interaction are collected in Appendix~\ref{appendix1}. The numerical results are presented in Table~\ref{Table-saturation}. The values of $L_{2k}$ 
provide a good handle on the density dependence of the corresponding $E_{\mathrm{sym}, 2k} (\rho)$ contributions. At second order, the slope 
parameter $L$ is positive in all the interactions. It goes from $L=17.57$ MeV in D260 to $46.53$ MeV in D280.
This large variation of the $L$ value indicates that the density dependence of the symmetry energy is poorly constrained with these forces \cite{PRC90SellahewaArnauRios2014}. 
We also emphasize that all forces in Table \ref{Table-saturation} have a low slope parameter, under $50$ MeV, and thus correspond to soft symmetry energies \citep{Tsang2012,Lattimer2013,Li:2013ola,Vinas:2013hua,Roca-Maza:2015eza,Lattimer2016}.
Indeed, we see that the $L$ values in Table \ref{Table-saturation} are below or on the 
low side of recent results proposed from microscopic calculations, such as 
$L=43.8$--48.6 MeV \cite{Birkhan:2016qkr}, $L=20$--65 MeV \cite{PRC95Holt2017} and 
$L=45$--70 MeV \cite{Drischler1710.08220}.

The higher-order slope parameters $L_4$ and $L_6$ are in keeping with the density dependence of $E_{\mathrm{sym}, 4}$ and $E_{\mathrm{sym}, 6}$, respectively.
As expected for two quantities that are difficult to constrain with finite nuclei properties, there are large 
systematic variations of the values of both $L_4$ and $L_6$. $L_4$ goes from about $-2$ MeV (D1N) to 
$4.7$ MeV (D260) and $L_{6}$ is in the range of $0.1-1.2$ MeV for the different forces. 
Interestingly, we find a one-to-one correspondence between group 1 and group 2 forces and the sign of $L_4$. For group 1 forces, $E_{\mathrm{sym}, 4} (\rho)$ has already reached a maximum at saturation density and tends to decrease with density (cf.~Fig.~\ref{fig:esym4});
consequently, $L_4$ is negative. On the contrary, group 2 forces have positive $L_4$, reflecting the increasing nature 
of $E_{\mathrm{sym}, 4} (\rho)$ with density. In contrast to $L_4$, 
the values of $L_{6}$ are always positive. This is a reflection 
of the fact that the maximum of $E_{\mathrm{sym}, 6} (\rho)$ occurs somewhat above saturation density, as shown in Fig.~\ref{fig:esym6}. 
It is worth noting that in absolute terms the value of the $L_{2k}$ parameters decreases with increasing order of the expansion, i.e., we have $|L_6| < |L_4| < |L|$. This indicates that the dominant density dependence of the isovector part of the 
functional is accounted for by the second-order parameter $L$.

\subsection{Parabolic approximation}
\label{sec:PA}

A parabolic approximation (PA) has been widely used in the literature to evaluate 
the energy of asymmetric matter with isospin asymmetry $\delta$ by interpolation of 
the energies in symmetric matter and in pure neutron matter, i.e.,
\begin{equation}\label{enerPA}
 E_b (\rho, \delta) = E_b (\rho, \delta=0) (1-\delta^2) \,+\, E_b (\rho, \delta=1) \delta^2 \, .
\end{equation}
In this case, the symmetry energy coefficient, which we will denote as 
$E_{\mathrm{sym}}^{PA}(\rho)$ in the following, is given by the difference between the 
energy per particle in pure neutron matter and in symmetric nuclear matter:
\begin{equation}\label{PA}
 E_{\mathrm{sym}}^{PA} (\rho) = E_b (\rho, \delta=1)-E_b (\rho, \delta=0) . 
\end{equation} 
This expression is often used in microscopic approaches, where calculations of asymmetric matter and its derivatives 
are not necessarily straightforward \cite{Baldo1999,Vidana2009}.
With the Taylor expansion in Eq.~(\ref{EoS}) taken up to order $\delta^2$ and setting $\delta=1$, one finds 
$E_{\mathrm{sym}}^{PA} (\rho) = E_{\mathrm{sym}, 2}(\rho)$. However, 
$E_{\mathrm{sym}}^{PA} (\rho)$ includes, 
in principle, contributions from all $E_{\mathrm{sym}, 2k}(\rho)$ terms:
\begin{equation}\label{eq:PA_2k}
 E_{\mathrm{sym}}^{PA} (\rho) = \sum_k E_{\mathrm{sym}, 2k} (\rho) \, .
\end{equation} 
Large values of the higher-order corrections to $E_{\mathrm{sym}, 2}(\rho)$ will spoil the correspondence between the two quantities. 

\begin{figure}[t!]
 \centering
 \includegraphics[width=1\linewidth, clip=true]{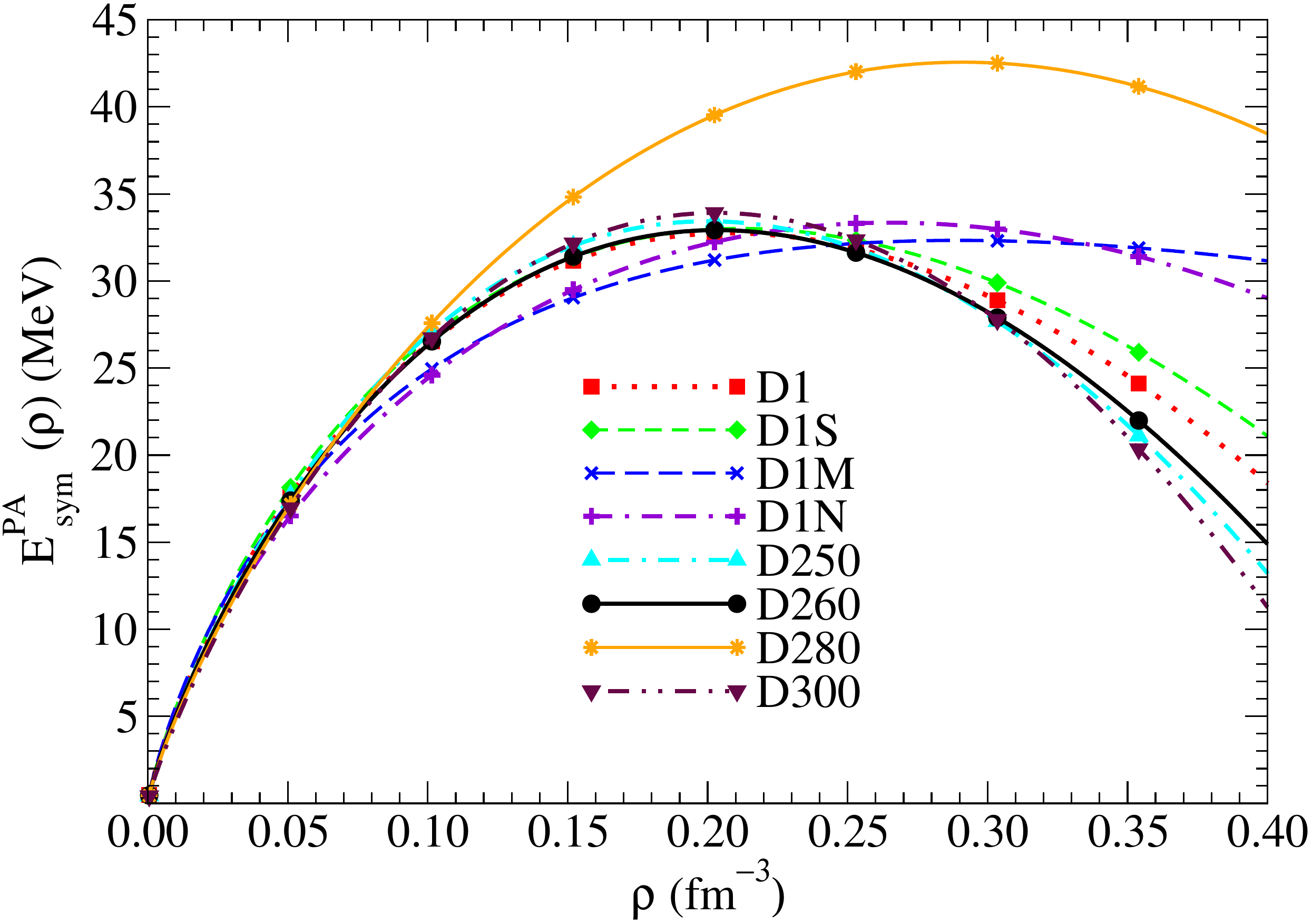}
  \caption{Density dependence of the symmetry energy coefficient in the parabolic approximation [Eq.~(\ref{PA})] for different Gogny interactions.}
\label{fig:esymPA}
\end{figure}

In Fig.~\ref{fig:esymPA} we show the results for $E_\mathrm{sym}^{PA}(\rho)$ from the different Gogny
functionals. We find a similar picture to that of Fig.~\ref{fig:esym2}.  
At subsaturation densities, the symmetry energies $E_{\mathrm{sym}}^{PA}(\rho)$ of all the forces are quite close to each other. At and above saturation, there are markedly different behaviors. 
Usually, $E_{\mathrm{sym}}^{PA}(\rho)$ reaches a maximum and then starts to decrease up to a given density 
where it becomes negative. 

\begin{figure}[t!]
 \centering
 \includegraphics[width=1\linewidth, clip=true]{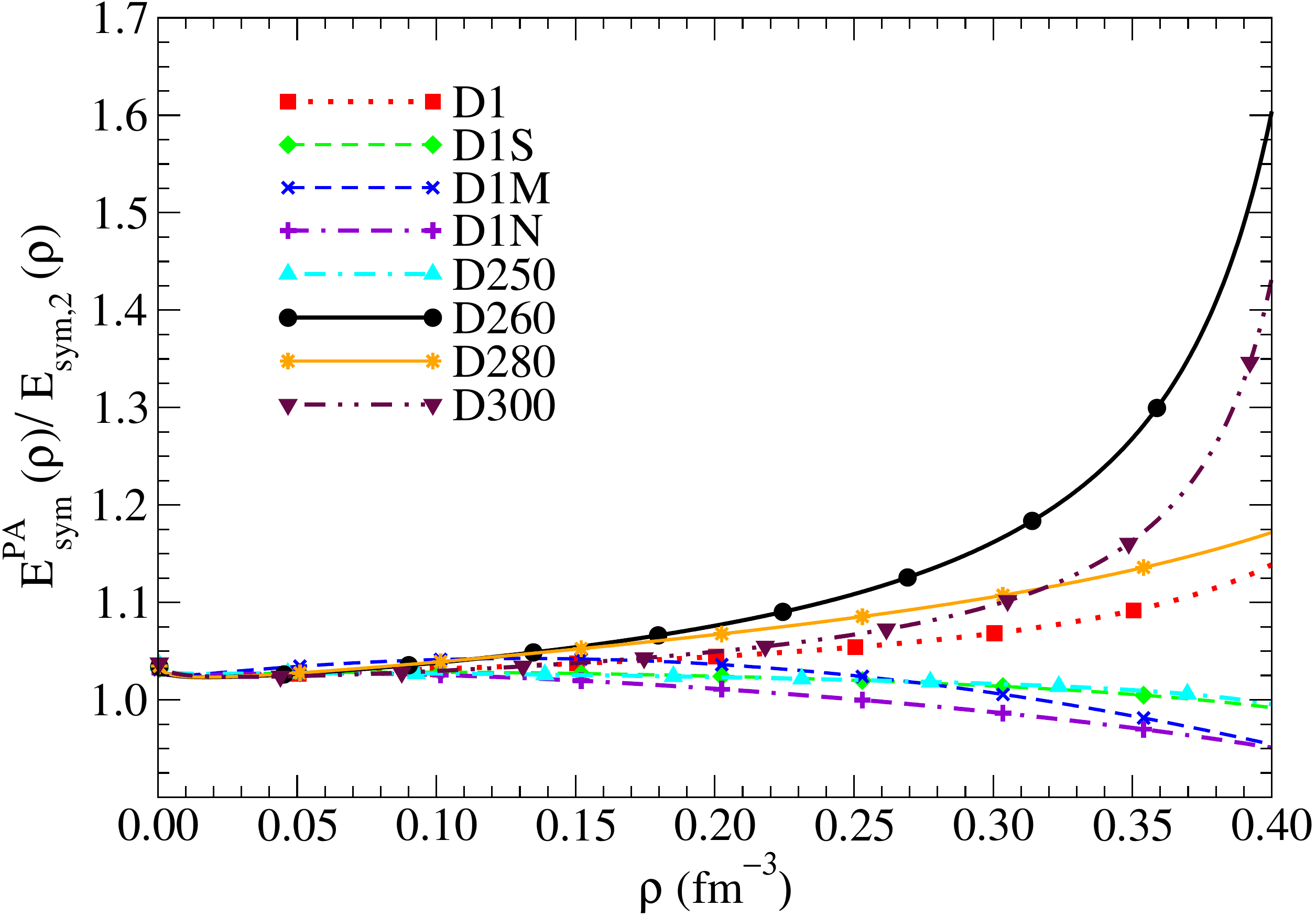}
  \caption{Density dependence of the ratio $E_{\mathrm{sym}}^{PA} (\rho)/E_{\mathrm{sym}, 2} (\rho)$ for different Gogny interactions.}
\label{fig:esymPAesym2}
\end{figure}

In order to analyze better the differences between $E_\mathrm{sym}^{PA}(\rho)$ and $E_{\mathrm{sym}, 2}(\rho)$,
we plot in Fig.~\ref{fig:esymPAesym2} the ratio $E_{\mathrm{sym}}^{PA}(\rho)/E_{\mathrm{sym}, 2}(\rho)$.  
At low densities $\rho \lesssim 0.1$ fm$^{-3}$, the symmetry energy calculated with the parabolic law is always a little
larger than calculated with Eq.~(\ref{esym}) for $k=1$. The ratio is approximately 1.025 irrespective of the functional. 
This is relatively consistent with the zero-density limit of a free Fermi gas,  which has a ratio 
$E_{\mathrm{sym}}^{PA}(\rho)/E_{\mathrm{sym}, 2}(\rho) = \frac{9}{5}(2^{2/3}-1) \approx 1.06$. 
At densities $\rho \gtrsim 0.1$ fm$^{-3}$, the ratios change depending on the Gogny force. 
Here, group~1 and group~2 parametrizations again show two distinct behaviors. In group~1 (D1S, D1M, D1N, D250), 
the ratio becomes smaller than $1$ at large densities, whereas in group~2 (D1, D260, D280, D300), it increases with density.
There is a clear resemblance between Fig.~\ref{fig:esymPAesym2} and the top panel of Fig.~\ref{fig:E46E2_Gogny}. 
Indeed, Eq.~(\ref{eq:PA_2k}) suggests that the two ratios are connected,
\begin{equation}
\frac{ E_{\mathrm{sym}}^{PA} (\rho) }{ E_{\mathrm{sym},2} (\rho) } = 1 + \frac{E_{\mathrm{sym}, 4} (\rho) }{E_{\mathrm{sym}, 2} (\rho) } + \cdots \, ,
\end{equation} 
as long as the next-order term $\frac{E_{\mathrm{sym}, 6} (\rho) }{E_{\mathrm{sym}, 2} (\rho) }$ is small. The behavior of the ratio $\frac{ E_{\mathrm{sym}}^{PA} (\rho) }{ E_{\mathrm{sym},2} (\rho) }$ can therefore be discussed in similar terms as the ratios shown in Fig.~\ref{fig:E46E2_Gogny}. 
As discussed earlier in the context of Eq.~(\ref{eq:esym46}), $E_{\mathrm{sym}, 4}(\rho)$ and $E_{\mathrm{sym}, 6}(\rho)$ are entirely determined by the exchange contributions that are proportional to the constants ${\cal C}_i$ and ${\cal D}_i$ and the functions $G_n (\mu_i k_F)$.

We include in Table \ref{Table-saturation} the results for $E_\mathrm{sym}^{PA}(\rho_0)$ at  
saturation density for the Gogny functionals. The values are of approximately 30--36 MeV.
In general, $E_{\mathrm{sym}}^{PA}(\rho_0)$ is larger than $E_{\mathrm{sym}, 2}(\rho_0)$ for the same interaction.
This is in accordance with Eq.~(\ref{eq:PA_2k}) and the fact that both $E_{\mathrm{sym}, 4}(\rho_0)$ and $E_{\mathrm{sym}, 6}(\rho_0)$ are positive (Table~\ref{Table-saturation}). 
When these are added up to the value of $E_{\mathrm{sym,2}}(\rho_0)$, one finds a very close agreement with $E_{\mathrm{sym}}^{PA}(\rho_0)$. The differences, about a percent, should be explained in terms of relatively small $k>3$ contributions.

Moreover, analogously to the definition of Eq.~(\ref{eq:L2k}), the slope parameter using the PA can be computed as 
\begin{eqnarray}
   L_{PA}&=& \left.3 \rho_0 \frac{\partial E_{\mathrm{sym}}^{PA} (\rho)}{\partial \rho} \right|_{\rho_0} .
\end{eqnarray}
The $L_{PA}$ values are displayed in the last row of Table~\ref{Table-saturation}.
There are again differences between the two groups of functionals. 
In group 1 forces, such as D1S, D1M, D1N, or D250, the $L_{PA}$ values are fairly close to the slope parameter $L$. In contrast, group 2 forces have $L_{PA}$ values that are substantially larger than $L$. For example, the relative 
differences between $L_{PA}$ and $L$ are of the order of $40 \%$ for D260 and $15 \%$ for D280. This again may be explained in terms of the higher-order $L_{2k}$ 
contributions, which add up to give $L_{PA}$ analogously to Eq.~(\ref{eq:PA_2k}).

This points to an important conclusion of this paper. For Gogny interactions, the parabolic approximation seems 
to work relatively well at the level of the symmetry energy. For the slope parameter, however, the contribution of 
$L_4$ can be large and spoil the agreement between the approximated $L_{PA}$ and $L$. 
$L_4$ is a density derivative of $E_\text{sym,4}$, which, as shown in Eq.~(\ref{eq:esym46}), is entirely determined by the exchange finite-range terms in the Gogny force.
The large values of 
$L_4$ are therefore due to isovector finite-range exchange contributions. We therefore conclude that exchange contributions play 
a very important role in the slope parameter. These terms can provide substantial (in some cases of order $30 \%$) corrections and should be explicitly considered when it is possible to do so \cite{Vidana2009}. 

\section{Neutron star core-crust transition}
\label{sec:corecrust}

\subsection{$\beta$-stable neutron star matter}
\label{subsec:betastable}
In $\beta$-stable \textit{npe} matter, the URCA reactions
\begin{align}
 n \rightarrow p + e^- + \bar{\nu}_e \qquad
 p+ e^- \rightarrow n + \nu_e 
\end{align}
take place simultaneously. Assuming that the neutrinos leave the system, $\beta$-equilibrium leads to the condition 
\begin{equation}\label{betaeq}
 \mu_{np} \equiv \mu_n - \mu_p = \mu_e ,
\end{equation}
where $\mu_n$, $\mu_p$, and $\mu_e$ are the chemical potentials of neutrons, protons, and electrons, respectively.
The analytical expression of the nucleonic chemical potentials for the Gogny interaction is provided in Appendix~\ref{appendix_p}, whereas the electronic chemical potential is given by Eq.~(\ref{eq:potentialele}) with $\rho_e = \rho_p$ due to charge neutrality. Ultimately, the condition of Eq.~(\ref{betaeq}) is an implicit equation for the isospin asymmetry $\delta$ that at each baryon density $\rho$ allows the system to be $\beta$-equilibrated.
 
Recalling Eq.~(\ref{eq:potentials}) for the neutron and proton chemical potentials, the $\beta$-equilibrium condition can be written as
\begin{eqnarray}\label{betamatter-full}
2 \frac{\partial E_b(\rho,\delta)}{\partial \delta} = \mu_e,
\end{eqnarray}
where $E_b(\rho,\delta)$ is the baryon energy per particle.
Now, if we replace in Eq.~(\ref{betamatter-full}) the full expression for $E_b(\rho,\delta)$ with its 
Taylor expansion in powers of $\delta^2$, given by Eq.~(\ref{EoS}), the $\beta$-equilibrium condition becomes
\begin{eqnarray}\label{betamatter}
&& 4 \delta E_{\mathrm{sym}, 2}(\rho) + 8 \delta^3 E_{\mathrm{sym}, 4}(\rho) \nonumber
 \\
 && \mbox{} + 12 \delta^5 E_{\mathrm{sym}, 6}(\rho) + \mathcal{O}(\delta^{7}) = \mu_e \, .
\end{eqnarray}
Upon using the PA discussed in Sec.~\ref{sec:PA},
it is easy to see that the $\beta$-equilibrium condition takes the form
\begin{eqnarray}\label{betamatter-PA}
4 \delta E_{\mathrm{sym}}^{PA}(\rho) = \mu_e \, .
\end{eqnarray}

Employing the full EoS of the interaction, 
the solution of Eq.~(\ref{betaeq}) [or, equivalently, Eq.~(\ref{betamatter-full})] will
yield the exact isospin asymmetry of $\beta$-equilibrium for Gogny forces. 
Solving Eqs.~(\ref{betamatter}) and (\ref{betamatter-PA}) instead, we 
will be able to gauge the quality of replacing the exact isospin dependence of the 
interaction by the different approximations of the symmetry energy.

\begin{figure}[t!]
 \centering
 \includegraphics[width=1\linewidth, clip=true]{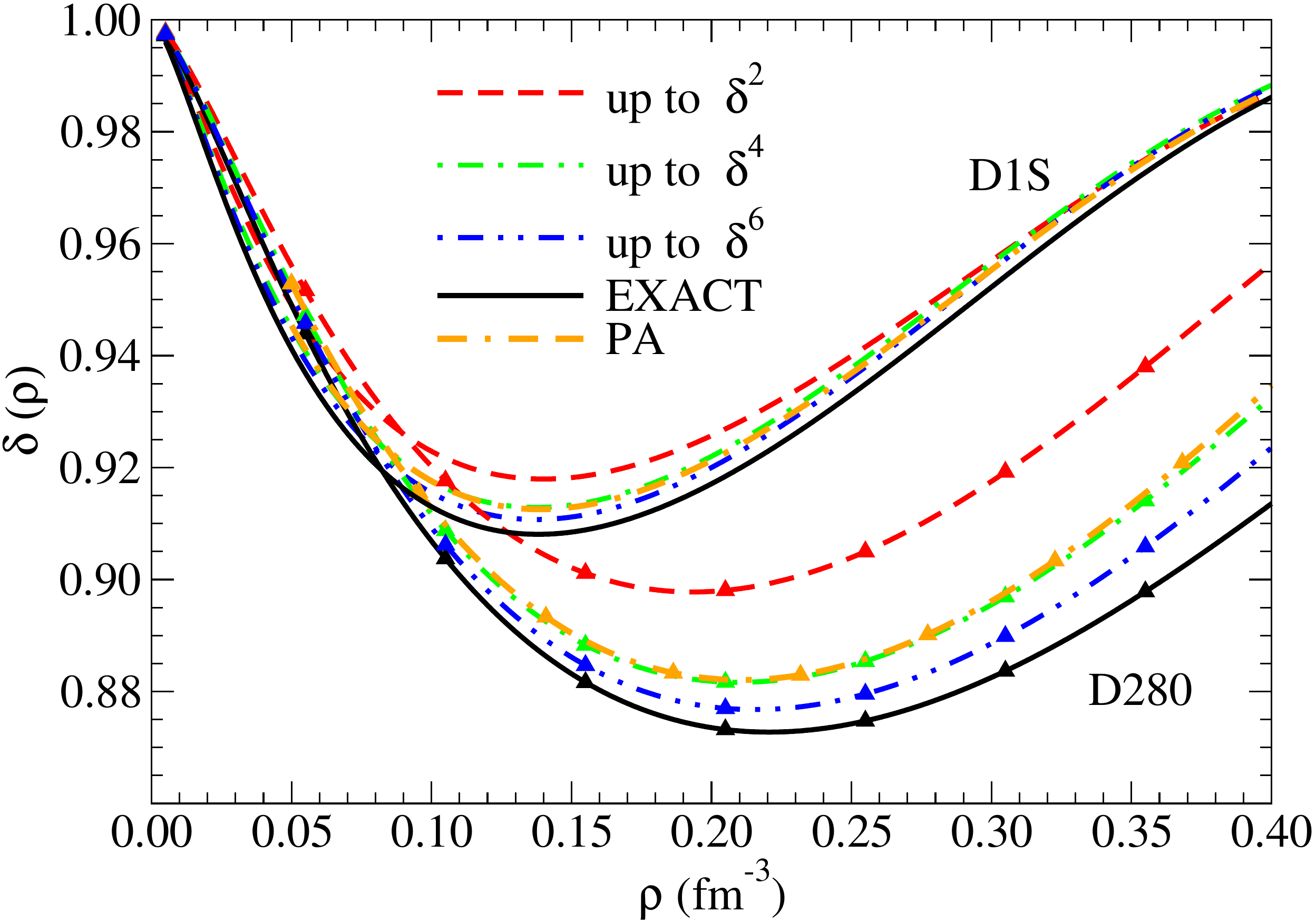}\\
  \caption{Density dependence of the isospin asymmetry in $\beta$-stable matter calculated using the exact expression of the EoS or the expression in Eq.~(\ref{EoS})
  up to second, fourth, and sixth order for the D1S and D280 interactions. The results of the parabolic approximation are also included.}
 \label{fig:delta}
\end{figure}

We present in Fig.~\ref{fig:delta} the asymmetry of $\beta$-stable matter as a function of density calculated using different approximations for two illustrative cases. Namely, we show the results for the 
D1S force (lines without symbols) that has a low slope parameter $L=22.4$~MeV and the results for D280 (lines with triangles) 
that has $L=46.5$~MeV, the largest $L$ value of the analyzed forces (cf.\ Table~\ref{Table-saturation}).
We include in Fig.~\ref{fig:delta} the results obtained with the exact EoS (black solid line), as well as those obtained with the expansion (\ref{EoS}) of the EoS up to second order (red dashed lines), fourth order (green dash-dotted lines), and sixth order (blue dash-double-dotted lines). We also provide results with the parabolic approximation (orange double-dash-dotted lines). 
In general, there is a trend, that in models with softer symmetry energy, like D1S,  
there is an overall larger isospin asymmetry at densities above $\sim 0.1$ fm$^{-3}$. 
In other words, the system is more neutron-rich at these densities for D1S as compared to D280. It is in consonance with the fact that for the same density range the symmetry energy of D1S is smaller than in D280, as can be seen in Fig.~\ref{fig:esym2}.
Importantly, we also find that, when one uses the Taylor expansion~(\ref{EoS}) of the EoS up to second order (i.e., $E_b(\rho, \delta) = E_b(\rho,0) + E_{\mathrm{sym}, 2} (\rho) \delta^2$), the predicted values for the $\beta$-equilibrium asymmetry are far from the results obtained with the exact EoS. 
The agreement improves as the approximations of the EoS increase in order, but even with the terms up to sixth order, the $\beta$-equilibrium asymmetries are not in line with the values of the exact EoS. 
As for the PA results (which correspond to using $E_b(\rho, \delta) = E_b(\rho,0) + E_{\mathrm{sym}}^{PA} (\rho) \delta^2$), it is interesting to note that they are significantly different from those obtained in the second-order approximation. In fact, for the functionals under consideration, the PA asymmetries are overall closer to the exact asymmetries than the second-order values. 

\begin{figure}[t!]
 \centering
 \includegraphics[width=1\linewidth, clip=true]{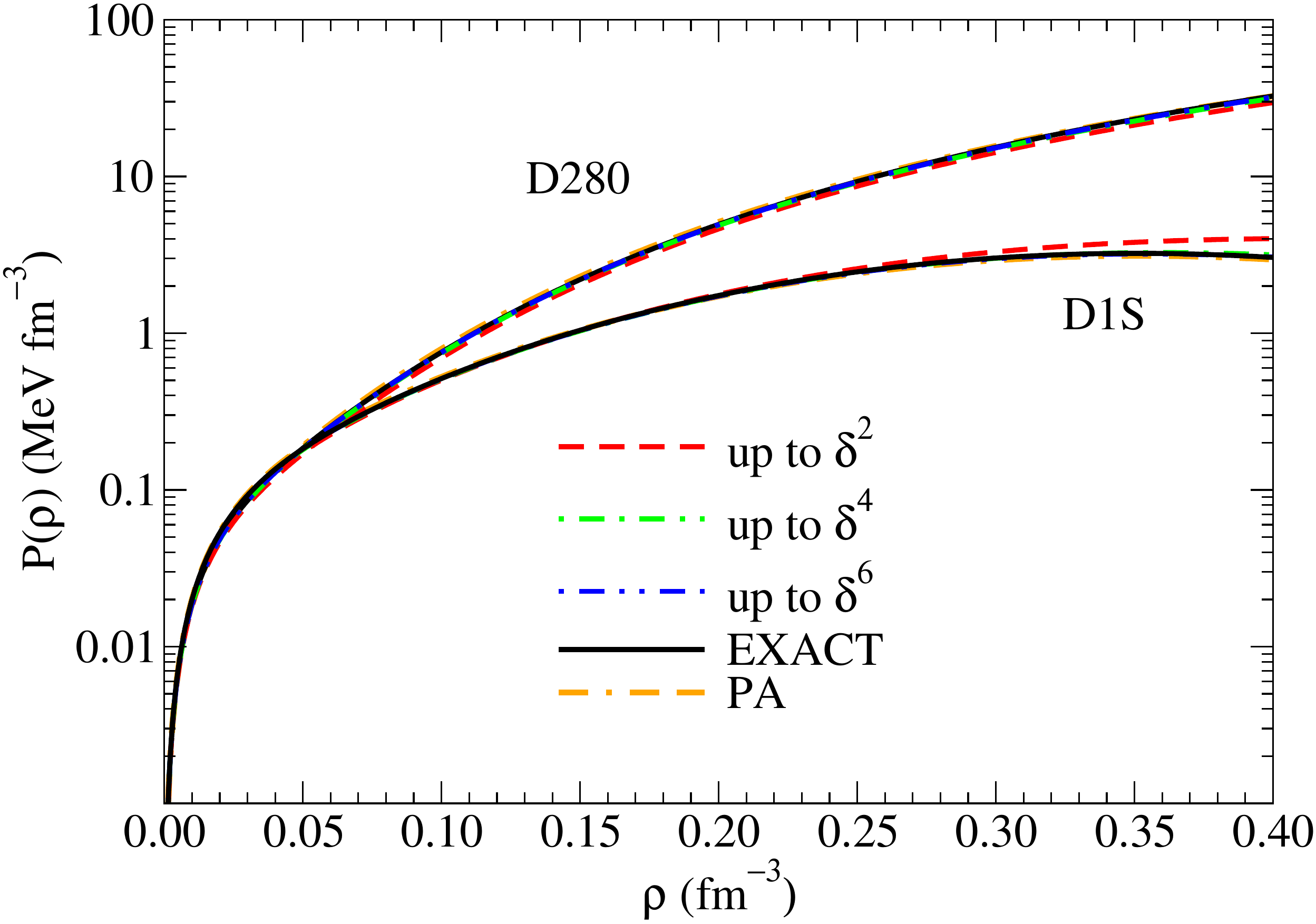}\\
  \caption{Density dependence of the pressure in $\beta$-stable matter calculated using the exact expression of the EoS or the expression in Eq.~(\ref{EoS})
  up to second, fourth, and sixth order for the D1S and D280 interactions. The results of the parabolic approximation are also included. The vertical axis is in logarithmic scale.}
 \label{fig:pressure}
\end{figure}

We display in Fig.~\ref{fig:pressure} the pressure of $\beta$-stable matter, including the (small) electron contribution, 
for the same Gogny forces. We show results calculated using the exact $\beta$-equilibrium condition, as well as the different 
approximations and the PA. For the D1S force, the relative differences between the pressure calculated at second 
order and the pressure of the exact EoS are of $30 \%$ at the largest density ($0.4$ fm$^{-3}$) of the figure. With the corrections up to sixth order included, the differences reduce to $1 \%$. For D280, these differences are of $10 \%$ and $1.5 \%$, respectively.
In all cases, adding more terms in the expansion brings the results closer to the pressure of the exact EoS. The results for the pressure are in keeping with the pure neutron matter predictions of Ref.~\cite{PRC90SellahewaArnauRios2014} 
and the $\beta$-stable calculations of Ref.~\cite{Loan2011}.

\subsection{Core-crust transition from the thermodynamical method}
\label{sec:thermodynamicalmethod}

In order to predict the transition between the core and the crust of the neutron star, we apply the so-called thermodynamical method \cite{PRC70Kubis2004,PRC76Kubis2007}
which has been widely used in the literature \cite{AJ697Xu2009,PRC81Moustakidis2010,PRC85Cai2012,PRC86Moustakidis2012,PRC89Seif2014,TRRoutray}.
Within this approach, the stability of the neutron star core is discussed in terms of its bulk properties. 
The following mechanical and chemical stability conditions set the boundaries of the homogeneous core:
\begin{eqnarray}\label{cond1}
 -\left( \frac{\partial P}{\partial v} \right)_{\mu_{np}} & > & 0 ,
\\[2mm]
\label{cond2}
 -\left( \frac{\partial \mu_{np}}{\partial q} \right)_v & > & 0 .
\end{eqnarray}
Here, $P$ is the total pressure of $\beta$-stable matter, defined in Eq.~(\ref{eq:pressure}); $\mu_{np}$ is the difference between the neutron and proton chemical potentials [Eq.~(\ref{betaeq})];
$v=1/\rho$ is the volume per baryon; and $q$ is the charge per baryon. 
Calculations with the dynamical method \cite{NPA175BAYM1971,NPA584Pethick1995,AJ697Xu2009,NPA789Ducoin2007,PRC83Ducoin2011}, the RPA \cite{Horowitz:2000xj,Carriere:2002bx,Fattoyev:2010tb}, 
or with a quantum mechanical approach based on response functions \cite{Ducoin2008,DePace2016} can also be implemented, although the presence of the finite-range exchange term in the Gogny interaction is a non-trivial complication. The thermodynamical approach is the long-wavelength limit of the dynamical method and requires the convexity of the energy per particle in the single phase when neglecting the Coulomb interaction~\cite{AJ697Xu2009,PRC81Moustakidis2010}.

First, we consider the mechanical stability condition in Eq.~(\ref{cond1}). 
The electron pressure does not contribute to this term, due to the fact that
the derivative is performed at constant $\mu_{np}$. 
In $\beta$-equilibrium, this involves a constant electron chemical potential $\mu_e$ and, 
because the  electron pressure 
in Eq.~(\ref{eq:press_el}) is a function of $\mu_e$ only, the derivative of $P_e$ with respect to $v$ vanishes.
Equation~(\ref{cond1}) can therefore be rewritten as
\begin{equation}\label{cond1Pb}
 -\left( \frac{\partial P_b}{\partial v} \right)_{\mu_{np}} >0.
\end{equation}
Moreover, the isospin asymmetry of the $\beta$-stable system is a function of density, $\delta(\rho)$. 
With $\mu_{np} = 2 \partial E_b/\partial \delta$, and using Eq.~(\ref{eq:Pb}) for baryons,
we can express the mechanical stability condition as \cite{AJ697Xu2009, PRC85Cai2012,  PRC81Moustakidis2010, PRC86Moustakidis2012}
\begin{eqnarray}\label{cond11}
  -\left( \frac{\partial P_b}{\partial v} \right)_{\mu_{np}} &=&\rho^2 \left[ 2 \rho 
  \frac{\partial E_b (\rho, \delta)}{\partial \rho} + \rho^2 \frac{\partial^2 E_b (\rho , \delta)}{\partial \rho^2}  \right. \nonumber
  \\
 &&\left. -\frac{\left( \rho \frac{\partial^2 E_b (\rho , \delta)}{\partial \rho \partial \delta} \right)^2}{\frac{\partial^2 
 E_b (\rho , \delta)}{\partial \delta^2}}\right] >0 .
\end{eqnarray}
In the chemical stability condition of Eq.~(\ref{cond2}), the charge $q$ can be written as $q=x_p - \rho_e/\rho$, 
where \mbox{$x_p = (1- \delta)/2$} is the proton fraction. 
In the ultrarelativistic limit, the electron number density is related to the chemical potential by $\rho_e= \mu_e^3/(3\pi^2)$. 
We can thus recast (\ref{cond2}) as
\begin{equation}\label{cond22}
 -\left( \frac{\partial q}{\partial \mu_{np}} \right)_v= \frac{1}{4} \left[ \frac{\partial^2 E_b (\rho, \delta)}{\partial \delta^2}
 \right]^{-1} + \frac{\mu_e^2}{\pi^2 \rho}>0. 
\end{equation}
In the low-density regime of interest for the core-crust transition, the first term on the rhs is positive for the Gogny parametrizations studied here. With a second term that is also positive, we conclude that the inequality of Eq.~(\ref{cond22}) is fulfilled. 
Hence, the stability condition for $\beta$-stable matter can be expressed in terms of Eq.~(\ref{cond11}) alone, with the result 
\cite{PRC70Kubis2004, PRC76Kubis2007,AJ697Xu2009,PRC81Moustakidis2010}
\begin{align}\label{Vthermal}
  V_{\mathrm{ther}} (\rho) &= 2 \rho \frac{\partial E_b (\rho, \delta)}{\partial \rho} + \rho^2 
  \frac{\partial^2 E_b (\rho , \delta)}{\partial \rho^2}\nonumber
  \\
 &-
 \left( \rho \frac{\partial^2 E_b (\rho , \delta)}{\partial \rho \partial \delta} \right)^2 \left(\frac{\partial^2 
 E_b (\rho , \delta)}{\partial \delta^2} \right)^{-1}>0,
\end{align}
where we have introduced a thermodynamical potential, $V_\mathrm{ther} (\rho)$. 

If the condition for $V_\mathrm{ther} (\rho)$ is rewritten using the Taylor expansion of $E_b (\rho, \delta)$ given in Eq.~(\ref{EoS}), one finds
\begin{eqnarray}\label{eq:Vtherapprox}
 V_{\mathrm{ther}}  (\rho) &=& \rho^2 \frac{\partial^2 E_b (\rho, \delta=0)}{\partial \rho^2} + 2 \rho 
 \frac{\partial E_b (\rho, \delta=0)}{\partial \rho} \nonumber
 \\
&&+ \sum_{k} \delta^{2k} \left( \rho^2 \frac{\partial^2 E_{\mathrm{sym}, 2k}(\rho)}{\partial \rho^2} + 2 \rho 
\frac{\partial E_{\mathrm{sym}, 2k}(\rho)}{\partial \rho}\right) \nonumber
\\
&&-2\rho^2 \delta^2 \left( \sum_{k} k \delta^{2k-2}  \frac{\partial E_{\mathrm{sym}, 2k}(\rho)}{\partial \rho} \right)^2 \nonumber
\\
&&\times \left[ \sum_{k} (2k-1)k\delta^{2k-2} E_{\mathrm{sym}, 2k}(\rho)\right]^{-1} >0 . \nonumber 
\\
\end{eqnarray}
This equation can be solved order by order, together with the $\beta$-equilibrium condition, Eq.~(\ref{betamatter})
(or Eq.~(\ref{betamatter-PA}) in the PA case), to evaluate the influence on
the predictions for the core-crust transition of truncating the Taylor expansion of the EOS of asymmetric nuclear matter.
We collect in Appendix~\ref{appendix_thermal} the expressions for the derivatives of $E_b(\rho, \delta)$ that are needed to calculate $V_\mathrm{ther} (\rho)$ in Eqs.~(\ref{Vthermal}) and (\ref{eq:Vtherapprox}) for Gogny forces. 

\begin{figure}[t!]
 \centering
 \raggedright
 \includegraphics[width=1\linewidth, clip=true]{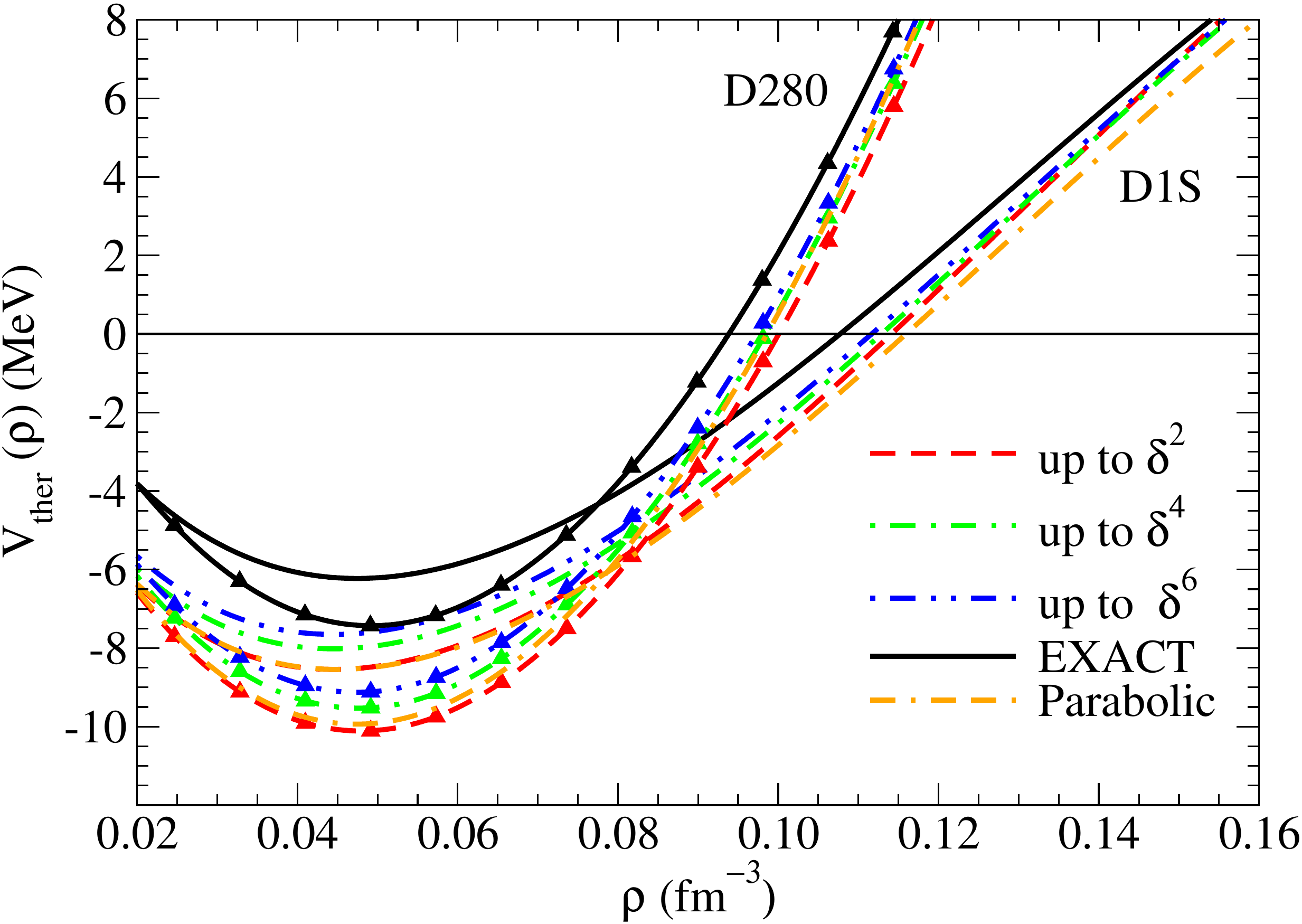}\\
  \caption{Density dependence of the thermodynamical potential $V_\mathrm{ther}(\rho)$ in $\beta$-stable matter 
  calculated using the exact expression of the EoS (solid lines) or the expression in Eq.~(\ref{EoS}) up to 
  second (dashed lines), fourth (dash-dotted lines) and sixth (dash-double-dotted lines) order for the D1S
  and D280 interactions. The results of the parabolic approximation are also included (double-dashed-dotted lines).}
 \label{fig:vthermal2}
\end{figure}

\begin{table*}[t!]
\begin{ruledtabular}
 \begin{tabular}{ccccccccc}
Force                       & D1     & D1S    & D1M    & D1N     & D250   & D260   & D280    & D300   \\ \hline

$\delta_t^{\delta^2}$       & 0.9215 & 0.9199 & 0.9366 & 0.9373  & 0.9167 & 0.9227 & 0.9202  & 0.9190 \\
$\delta_t^{\delta^4}$       & 0.9148 & 0.9148 & 0.9290 & 0.9336  & 0.9119 & 0.9136 & 0.9127  & 0.9128 \\
$\delta_t^{\delta^6}$       & 0.9127 & 0.9129 & 0.9265 & 0.9321  & 0.9101 & 0.9112 & 0.9110  & 0.9110 \\
$\delta_t^{\mathrm{exact}}$ & 0.9106 & 0.9111 & 0.9241 & 0.9310  & 0.9086 & 0.9092 & 0.9110  & 0.9096 \\
$\delta_t^{PA}$             & 0.9152 & 0.9142 & 0.9296 & 0.9327  & 0.9111 & 0.9153 & 0.9136  & 0.9134 \\ \hline

$\rho_t^{\delta^2}$         & 0.1243 & 0.1141 & 0.1061 & 0.1008  & 0.1156 & 0.1228 & 0.1001  & 0.1161 \\
$\rho_t^{\delta^4}$         & 0.1222 & 0.1129 & 0.1061 & 0.0996  & 0.1143 & 0.1198 & 0.0984  & 0.1145 \\
$\rho_t^{\delta^6}$         & 0.1211 & 0.1117 & 0.1053 & 0.0984  & 0.1131 & 0.1188 & 0.0973  & 0.1136 \\
$\rho_t^{\mathrm{exact}}$   & 0.1176 & 0.1077 & 0.1027 & 0.0942 & 0.1097 & 0.1159 & 0.0938 & 0.1109 \\
$\rho_t^{PA}$               & 0.1222 & 0.1160 & 0.1078 & 0.1027  & 0.1168 & 0.1171 & 0.0986  & 0.1142 \\ \hline

$P_t^{\delta^2}$            & 0.6279 & 0.6316 & 0.3326 & 0.4882  & 0.7034 & 0.5892 & 0.6984  & 0.6776 \\
$P_t^{\delta^4}$            & 0.6479 & 0.6239 & 0.3531 & 0.4676  & 0.6908 & 0.6483 & 0.7170  & 0.6998 \\
$P_t^{\delta^6}$            & 0.6452 & 0.6156 & 0.3554 & 0.4582  & 0.6811 & 0.6509 & 0.7053  & 0.6955 \\
$P_t^{\mathrm{exact}}$      & 0.6184 & 0.5817 & 0.3390 & 0.4164  & 0.6464 & 0.6272 & 0.6493  & 0.6647 \\
$P_t^{PA}$                  & 0.6853 & 0.6725 & 0.3986 & 0.5173  & 0.7368 & 0.6809 & 0.7668  & 0.7356 
\end{tabular}
\end{ruledtabular}
\caption{Values of the core-crust transition density $\rho_t$ (in fm$^{-3}$)
calculated using the exact expression of the EoS ($\rho_t^\mathrm{exact}$), the parabolic approximation ($\rho_t^{PA}$), 
or the approximations of the full EoS with Eq.~(\ref{EoS}) up to second ($\rho_t^{\delta^2}$), fourth ($\rho_t^{\delta^4}$) and sixth ($\rho_t^{\delta^6}$) order.
The table includes the corresponding values of the transition pressure $P_t$ (in MeV fm$^{-3}$) and isospin asymmetry~$\delta_t$.}
\label{Table-transition}
\end{table*}

We show in Fig.~\ref{fig:vthermal2} the density dependence of $V_\mathrm{ther} (\rho)$ 
in $\beta$-stable matter, calculated with the exact expression of the EoS (solid 
lines), with its Taylor expansion up to second, fourth and sixth order, and with the 
PA. An instability region characterised by negative $V_\mathrm{ther} (\rho)$ is found 
below $\rho \approx 0.09-0.11$ fm$^{-3}$. The condition $V_\mathrm{ther}(\rho_t) = 0$ 
defines the density $\rho_t$ of the transition from the homogeneous core to the crust.
We see in Fig.~\ref{fig:vthermal2} that adding more terms to Eq.~(\ref{EoS}) brings the 
results for $V_\mathrm{ther} (\rho)$ closer to the exact values. 
At densities near the core-crust transition, the higher-order results are rather 
similar, but differ significantly from the exact ones.
We note that, all in all, the order-by-order convergence of the $\delta^2$ expansion in 
$V_\mathrm{ther} (\rho)$ is slow. This indicates that the non-trivial isospin and 
density dependence arising from exchange terms needs to be considered in a complete 
manner for realistic core-crust transition physics 
\cite{PRC80Chen2009,Vidana2009,PRC89Seif2014}.
If we look at the unstable low-density zone, both the exact and the approximated 
results for $V_\mathrm{ther} (\rho)$ go to zero for vanishing density, but they keep a 
different slope. In this case, we have found that the discrepancies are largely 
explained by the differences in the low-density behavior of the approximated 
kinetic energy terms, in consonance with the findings of Ref.~\cite{TRRoutray}.

We next analyze more closely the properties of the core-crust transition, using both exact and order-by-order predictions. The complete results for the eight Gogny functionals are provided in numerical form in Table~\ref{Table-transition}. For a better understanding, we discuss each one of the key physical properties of the transition (asymmetry, density, and pressure) in separate figures. We plot our predictions as a function of the slope parameter $L$ of each functional, which does not necessarily provide a stringent correlation with core-crust properties \cite{PRC83Ducoin2011}. 
The slope parameter, however, can be constrained in terrestrial experiments and astrophysical observations \citep{Tsang2012,Lattimer2013,Lattimer2016,Li:2013ola,Vinas:2013hua,Roca-Maza:2015eza}
and is therefore an informative parameter in terms of the isovector properties of the functional.

\begin{figure}[t!]
 \centering
 \includegraphics[width=1\linewidth, clip=true]{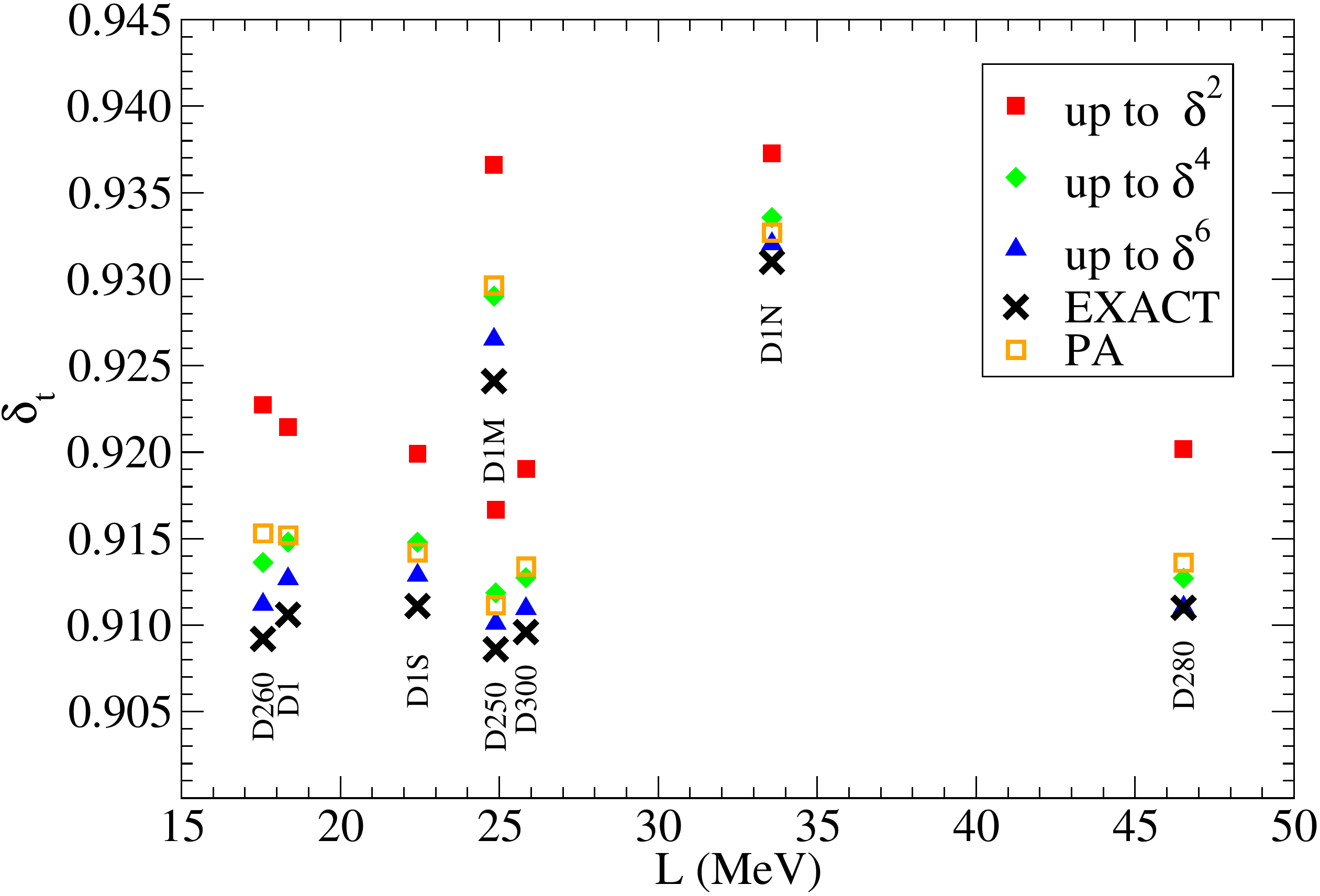}\\
   \caption{Core-crust transition asymmetry, $\delta_t$, as a function of the slope parameter $L$ calculated using the exact expression of the EoS (crosses), 
   and the approximations up to second (solid squares), fourth (solid diamonds) and sixth order (solid triangles). The parabolic approximation is also included (empty squares).}
 \label{fig:deltat}
\end{figure}

In Fig.~\ref{fig:deltat}, we display the results for the transition asymmetry, $\delta_t$. Black crosses correspond to the calculations with the exact EoS. We find that the Gogny forces predict a range $0.909 \lesssim \delta_t \lesssim 0.931$ for the asymmetry at the transition point, or, in other words, proton fractions in the range $3.45\%\lesssim x_p \lesssim 4.55\%$. 
The D1N and D1M forces provide distinctively large transition asymmetries,
whereas the other interactions predict very similar values $\delta_t \approx 0.91$ in spite of having different slope parameters.
When using the Taylor expansion of the EoS up to second order (shown by red squares in the figure), 
the predictions 
for $\delta_t$ are, in all forces, well above the exact result. 
The fourth-order values (green diamonds) are still above the exact ones but closer, and the sixth-order calculations (blue triangles) produce results that are very close to the exact $\delta_t$. 
The $\delta_t$ values obtained with the PA (empty orange squares) differ from the second-order approximation and turn out to be closer to the exact results.

\begin{figure}[t!]
 \centering
 \includegraphics[width=1\linewidth, clip=true]{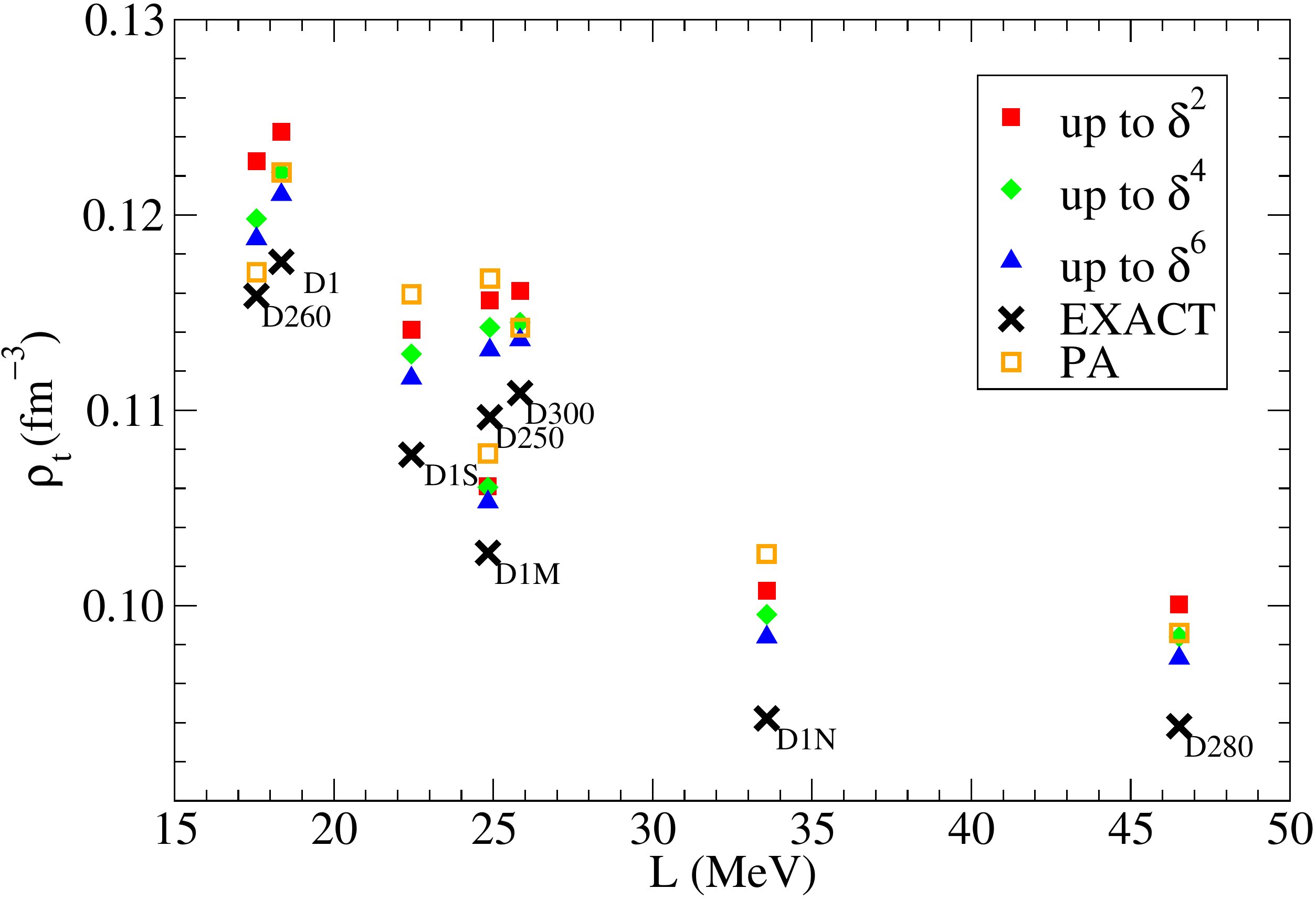}\\
  \caption{Core-crust transition density, $\rho_t$, as a function of the slope parameter $L$. Symbols are defined as in Fig.~\ref{fig:deltat}.}
 \label{fig:rhot}
\end{figure}

We show in Fig.~\ref{fig:rhot} the predictions for the density of the core-crust transition, $\rho_t$. The calculations with the exact EoS of the models give a window $0.094 \text{ fm}^{-3} \lesssim \rho_t \lesssim 0.118\text{ fm}^{-3}$. Again, we find that the approximations of the EoS only provide upper bounds to the exact values. 
The relative differences between the transition densities predicted using the $\delta^2$ approximation of the EoS
and the exact densities are about $4\%-7 \%$.
When the EoS up to $\delta^4$ is used, the differences are slightly reduced 
to $3\%-6 \%$. The sixth-order results remain at a similar level of accuracy, within $3\%-5 \%$. 
In other words, the order-by-order convergence for the transition density is very slow.
As mentioned earlier in the discussion of Fig.~\ref{fig:vthermal2}, the non-trivial density and isospin asymmetry dependence of the thermodynamical potential arising from the exchange contributions is likely to be the underlying cause of this slow convergence pattern. The results for $\rho_t$ of the PA do not exhibit a regular trend with respect to the other approximations. The PA estimate may happen to be closest to the exact $\rho_t$, as in D260, but it may also be the most distant, as in D1S and other models.    

Unlike the transition asymmetry $\delta_t$, we find that there is a decreasing quasi-linear correlation between the transition 
density $\rho_t$ and the slope parameter $L$. In fact, it is known from previous literature that the transition 
densities calculated with Skyrme interactions and RMF models have an anticorrelation with $L$ \cite{Horowitz:2000xj,AJ697Xu2009, PRC86Moustakidis2012,PRC83Ducoin2011,EPJAProvidencia2014,Fattoyev:2010tb,Pais2016}. We confirm 
this tendency and find that the transition densities calculated with Gogny functionals are in consonance with other 
mean-field models. Moreover, if we take into account the slope parameter of these interactions, the Gogny results are within the expected window of values provided by the Skyrme and RMF models \cite{AJ697Xu2009,Vidana2009}. 
The values of $\rho_t$ that we obtain are larger than some recent predictions \cite{Pais2016,Lim2017}, as expected from the relatively low values of $L$ of the Gogny forces. In the future, it might be interesting to explore Gogny parametrizations with larger values of $L$ to confirm this tendency.

\begin{figure}[b!]
 \centering
 \includegraphics[width=1\linewidth, clip=true]{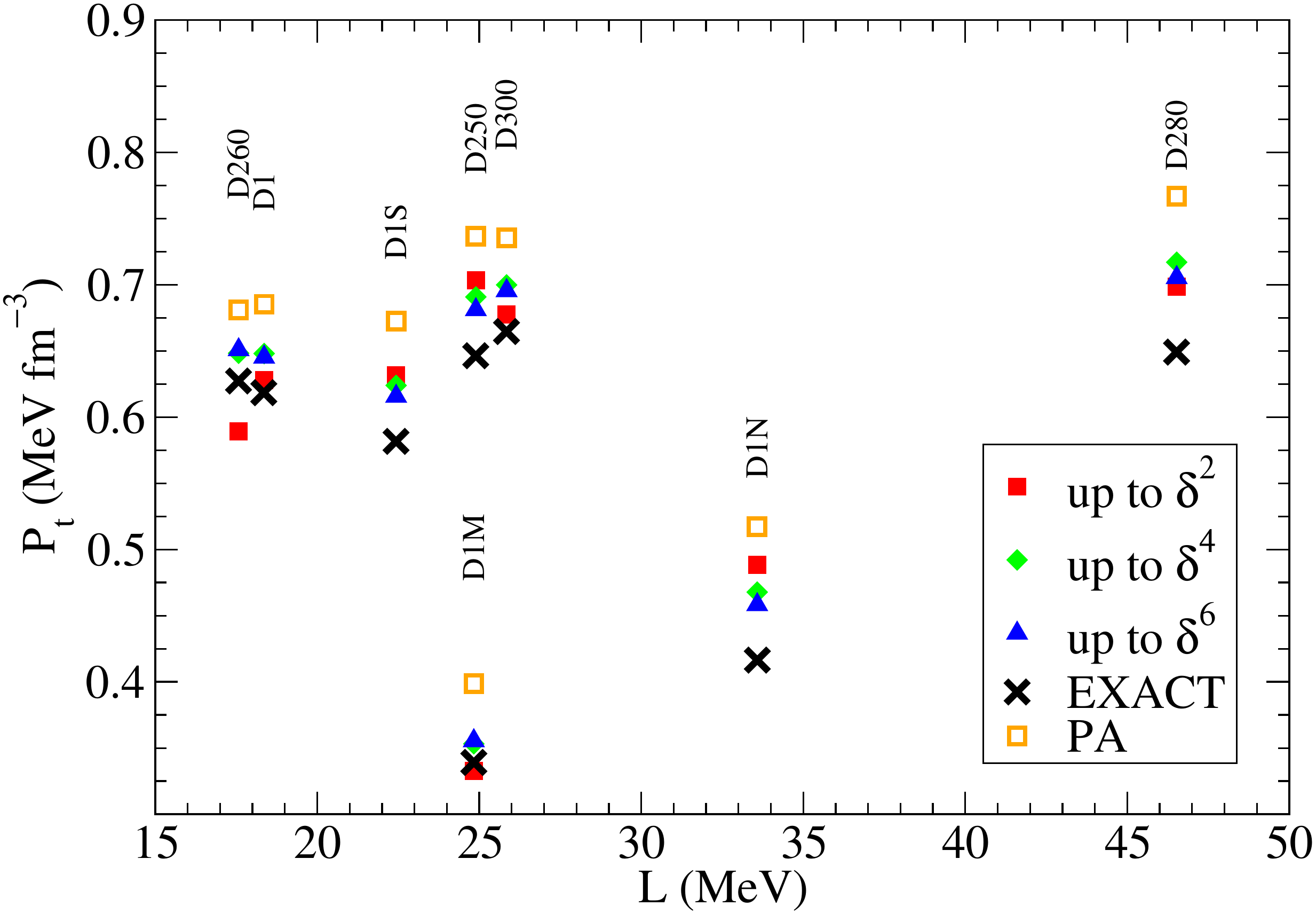}\\
   \caption{Core-crust transition pressure, $P_t$, as a function of the slope parameter $L$. Symbols are defined as in Fig.~\ref{fig:deltat}.}
 \label{fig:Pt}
\end{figure}

In Fig.~\ref{fig:Pt}, we present the pressure at the transition point, $P_t$, for the same interactions. The results of the exact Gogny EoSs lie in the range $0.339 \text{ MeV fm}^{-3} \lesssim P_t \lesssim 0.665 \text{ MeV fm}^{-3}$, with D1M and D1N giving the lower $P_t$ values. 
According to Ref.~\cite{AJ550LattimerPrakash2001}, in general, the transition pressure for realistic EoSs varies over a window $0.25 \text{ MeV fm}^{-3} \lesssim P_t \lesssim 0.65 \text{ MeV fm}^{-3}$. Gogny forces therefore seem to deliver reasonable predictions.
If we look at the accuracy of the isospin Taylor expansion of the EoS for predicting $P_t$, we find that the second-order approximation gives transition pressures above the values of the exact EoS in almost all of the forces. The differences are of about $2\%-17 \%$.
These differences become $3\%-12 \%$ at fourth order of the expansion, and $4\%-10 \%$ at sixth order. On the whole, Fig.~\ref{fig:Pt} shows that the order-by-order convergence for the transition pressure is not only slow, but actually erratic at times. 
For some parametrizations, like D1 or D300, the fourth- and sixth-order predictions for $P_t$ differ more from the exact value than if we stop at second order. 
We also see that the PA overestimates the transition pressure for all parametrizations---in fact, the PA provides worse predictions for the transition pressure than any of the finite-order approximations.

We note that we do not find a general trend with the slope parameter $L$ in our results for the pressure of the transition, i.e., Gogny forces with similar $L$ may have quite different pressure at the border between the core and the crust. As in the case of the transition density, the transition pressure has been studied in previous literature. However, the predictions on the correlation between the transition pressure and $L$ diverge \cite{AJ697Xu2009,PRC86Moustakidis2012,Fattoyev:2010tb,PRC90Piekarewicz2014,PRC81Moustakidis2010}.
In our case, we obtain that the transition pressure is uncorrelated with the slope parameter~$L$. 
The same was concluded in Ref.~\cite{Fattoyev:2010tb} in an analysis with RMF models.

\section{Neutron star structure} 
\label{sec:NS}

\subsection{Bulk properties of the stars}

With access to the analytical expressions for the pressure and the energy density in asymmetric matter, one can compute the mass-radius relation 
of neutron stars by integrating the TOV equations \cite{Shapiro1983,Glendenning2000,Haensel2007}. We have 
solved these equations for the above Gogny forces \cite{SellahewaPhD} using the $\beta$-equilibrium EoS with the exact isospin asymmetry dependence in the neutron star core.
Note that at high densities these conditions yield a pure neutron star with $\delta=1$, and we ignore the effects of an 
isospin instability at and beyond that point. At very low densities, we use the Haensel-Pichon EoS for the outer crust \cite{Douchin2001}. 
In the absence of microscopic calculations of the EoS of the inner crust with the Gogny forces, 
we adopt the prescription of previous works \cite{Carriere:2002bx,AJ697Xu2009,Zhang:2015vaa} by taking 
the EoS of the inner crust to be of the polytropic form $P=a+b\epsilon^{4/3}$, where $\epsilon$ denotes the mass-energy density.
The constants $a$ and $b$ are adjusted by demanding continuity at the inner-outer crust interface and at the core-crust transition point~\cite{Carriere:2002bx,AJ697Xu2009,Zhang:2015vaa}.
At the subsaturation densities of the inner crust, the pressure of matter is dominated by the 
relativistic degenerate electrons and a polytropic form with an index of average value of about $4/3$ is found
to be a good approximation to the EoS in this region \cite{Link1999,AJ550LattimerPrakash2001,Lattimer2016}.
For more accurate predictions of the crustal properties, it would be of great interest to determine the microscopic EOS of the inner crust with finite-range Gogny interactions \cite{Than2011}, which we leave for future work.

\begin{figure}[t!]
 \centering
 \includegraphics[width=1\linewidth, clip=true]{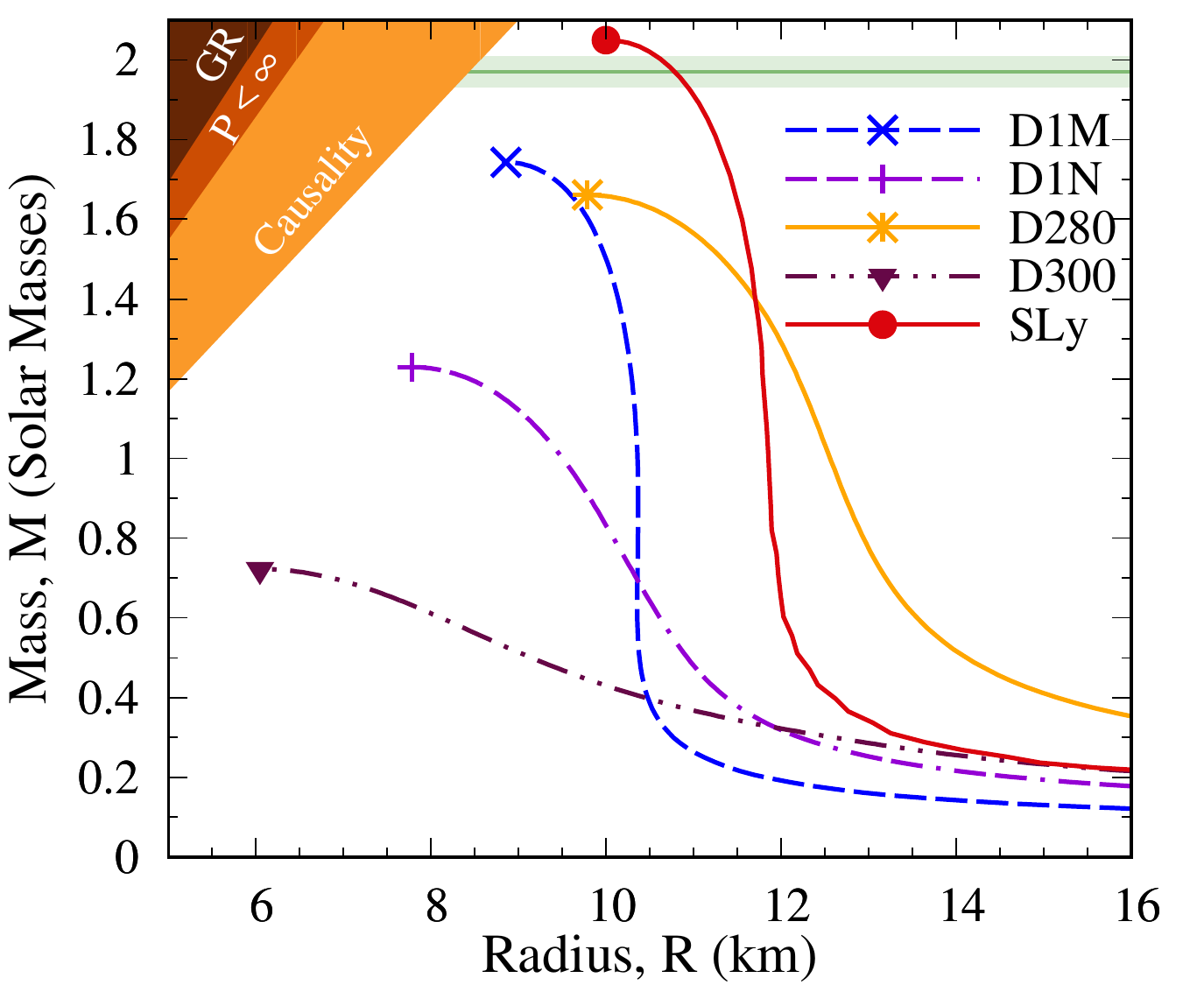}\\
  \caption{Mass-radius relation for the neutron stars produced with the four stable Gogny functionals and with the unified SLy EoS \cite{Douchin2001}. We show physically excluded regions in the upper-left corner as well as the accurate $M \approx 2 M_\odot$ mass measurement of Ref.~\cite{Demorest2010}. }
\label{fig:MR}
\end{figure}

The results for the mass-radius relationship are presented in Fig.~\ref{fig:MR}. We also show the predictions of the SLy unified neutron star EoS \cite{Douchin2001}, which we will use as a benchmark in our following discussions. 
It may be mentioned that we could have adopted other reference unified neutron star EoSs, like those recently developed from the Brussels-Montreal BSk models \cite{Potekhin:2013qqa} 
or from the Brueckner theory \cite{Baldo:2013ska,Sharma:2015bna}. 
First, we stress that Fig.~\ref{fig:MR} contains only the four Gogny functionals that provide numerically stable solutions for neutron stars. 
Second, and more important, all Gogny EoSs provide maximum neutron star masses that are well below the observational limit of $M \approx 2 M_\odot$ from Refs.~\cite{Demorest2010,Antoniadis2013}. 
As a matter of fact, only D1M and D280 are able to generate masses above the canonical $1.4 M_\odot$ value. 
The neutron star radii from these two EoSs are considerably different, however, with D1M producing stars with radii $R\approx 9-10.5$ km, and D280 stars with radii $R \approx 10-12$ km. 
These small radii for a canonical neutron star would be in line with recent extractions of stellar radii from quiescent low-mass x-ray binaries and x-ray burst sources, that have suggested values in the range of $9-13$ km \cite{Guillot:2014lla,Heinke:2014xaa,Ozel:2015fia,Lattimer2016,Ozel:2016oaf}. 
It appears that a certain degree of softness of the nuclear symmetry energy is necessary in order to reproduce small radii for a canonical mass neutron star~\cite{Chen:2015zpa,Jiang:2015bea,Tolos:2016hhl}. 
The parametrizations D1N and D300, in contrast to D1M and D280, generate neutron stars which are unrealistically small in terms of both mass and radius. 
One should of course be cautious in 
interpreting these results. Gogny forces have not been fit to reproduce high-density, neutron-rich systems and it is not surprising that some parametrizations do not yield realistic neutron stars. 

\begin{table}[t!]
\begin{ruledtabular}
\begin{tabular}{c | cccc}  
 & D1M & D280 & D1M & D280 \\
 & $M_\text{max}$ &  $M_\text{max}$ & $1.4 M_\odot$ & $1.4 M_\odot$ \\ \hline
$\rho_c$ (fm$^{-3}$) & 1.57 & 1.46 & 0.81 & 0.69  \\
$\epsilon_c$ ($10^{15}$ g cm$^{-3}$) & 3.65 & 3.28& 1.51 & 1.30 \\
$R$ (km) & 8.85 & 9.77 & 10.1 & 11.7 \\
$M$ ($M_\odot$) & 1.74 & 1.66 & 1.40 & 1.40 \\
$A$ ($10^{57}$) & 2.45 & 2.26 & 1.89 & 1.85 \\
$E_\text{bind}$ ($10^{53}$ erg) & 5.43 & 4.00 & 3.09 & 2.56 \\
$z_\text{surf}$ & 0.55 & 0.42 & 0.30 & 0.24 \\
$I$ ($10^{45}$ g cm$^{2}$) & 1.23 & 1.21 & 1.10 &1.27 
\end{tabular}
\end{ruledtabular}
\caption{Properties of the neutron star maximum mass and $1.4 M_\odot$ configurations for the D1M and D280 functionals. From top to bottom, we quote central number density, central mass-energy density, radius, mass, baryon number, binding energy, surface red shift, and moment of inertia of the star.}
\label{Table-NSs}
\end{table}

One could presumably improve these results by guaranteeing that, at least around the saturation region, the pressure of neutron-rich matter is compatible with neutron star observations \cite{Lattimer2013}. This could provide a Gogny force in the spirit of the well-known Skyrme SLy forces \cite{NPA627Chabanat1997,NPA635Chabanat1998}, which are still widely used in both nuclear structure and neutron star studies. For completeness, we provide data on the maximum mass and $1.4 M_\odot$ configurations of the neutron stars produced by D1M and D280 in Table~\ref{Table-NSs}. The maximum mass configurations are reached at central baryon number densities close to $\approx 10 \rho_0$, whereas $1.4 M_\odot$ neutron stars have central baryon densities close to around $4-5 \rho_0$. These large central density values are in keeping with the fact that the neutron matter Gogny EoSs are relatively soft, which require larger central densities to produce realistic neutron stars.

\begin{figure}[t!]
 \centering
 \includegraphics[width=1\linewidth, clip=true]{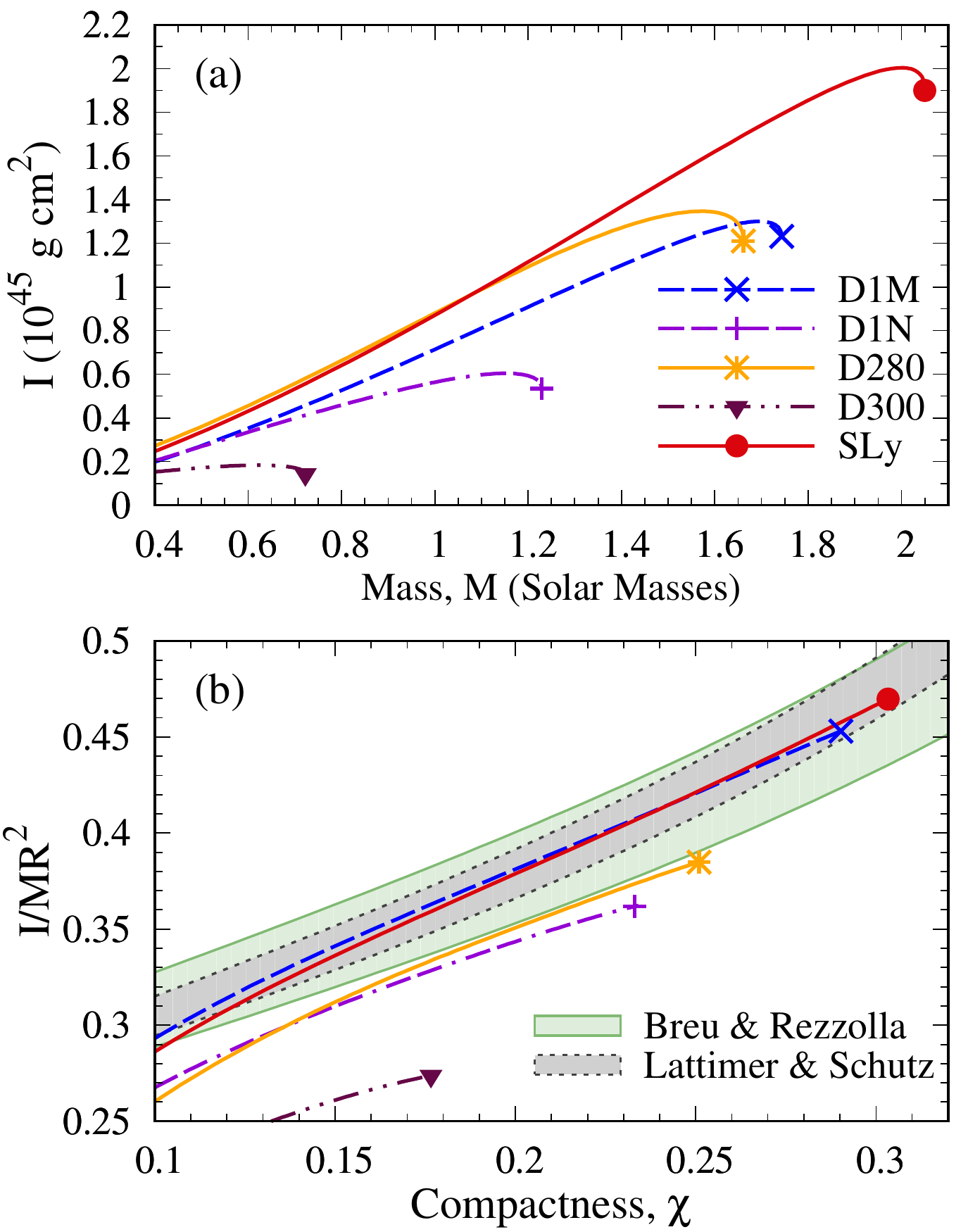}\\
   \caption{Panel (a): Neutron star moment of inertia as a function of mass for the four stable Gogny functionals and for the SLy EoS. Panel (b): Dimensionless ratio $I/MR^2$ as a function of compactness for the EoSs. The shaded area enclosed by a solid line is the parametrization provided by Breu and Rezzolla~\cite{Breu2016}. The shaded area enclosed by a dashed line is that of Lattimer and Schutz~\cite{Lattimer2005}. }
 \label{fig:momI}
\end{figure}

One property of interest, due to potential observational evidence in binaries as well as the connection to the core-crust transition, is the star's moment of inertia, $I$ \cite{Ravenhall1994,Haensel2007,Lattimer2005,Lattimer2016}.
To lowest order in angular velocity, the moment of inertia of the star can be computed from the static mass distribution and gravitational potentials encoded in the TOV equations \cite{Hartle1967}. We do not provide further details of the standard numerical procedure to obtain this quantity, but note that our code has been tested against the results of known EoSs \cite{Haensel2007,Ravenhall1994}. We show in panel (a) of Fig.~\ref{fig:momI} the results of the moments of inertia for the four Gogny parametrizations of interest. 
In agreement with the findings of the EoS, the moments of inertia are relatively small. In particular, we find that the moments of inertia of the Gogny parametrizations are below the predictions of SLy.  As expected, the maximum of $I$ is reached slightly below the maximum mass configuration for all forces \cite{Haensel2007}. 
In the case of the two most realistic EoSs (D1M and D280), we find a maximum value $I_\text{max} \approx 1.3-1.4 \times 10^{45}$ g cm$^2$. This is below the typical maximum values of $\approx 2 \times 10^{45}$ g cm$^2$ obtained with stiffer EoSs \cite{Haensel2007}. 
Our results for D1N are commensurate with those of Ref.~\cite{Loan2011}.

A useful comparison with the systematics of other neutron star EoSs is provided by the dimensionless quantity $\frac{I}{MR^2}$. This has been found to scale with the neutron star compactness which, in natural units, is
\begin{align}
\chi = \frac{GM}{R} .
\end{align}
In fact, in a relatively wide region of $\chi$ values, the dimensionless ratio $\frac{I}{MR^2}$ for the mass and radius combinations of several EoSs can be fitted by universal parametrizations \cite{Ravenhall1994,Lattimer2005,Breu2016}. We show in panel (b) of Fig.~\ref{fig:momI} this dimensionless ratio as a function of compactness for the four Gogny forces and the SLy EoS. Our results are compared to the recent fits from Breu and Rezzolla~\cite{Breu2016} (shaded area enclosed by a solid line) and the older results from Lattimer and Schutz~\cite{Lattimer2005} (shaded region enclosed by a dashed line).
 These fits have been obtained from a very wide range of different theoretical EoS predictions.
For compactness $\chi > 0.1$, only D1M falls within the wider range obtained with the parametrization of Breu and Rezzolla~\cite{Breu2016}. D280 is close to the lower limit of this fit, but well below the lower bounds of the fit in Lattimer and Schutz~\cite{Lattimer2005}. We find that, in spite of the significant differences in their absolute moments of inertia, both D1M and SLy produce dimensionless ratios which agree well with each other. In contrast, and as expected, D1N and, specially, D300 produce too small moments of inertia for a given mass and radius, and systematically fall below the fits. This again illustrates the inability of these two forces to create realistic neutron stars.

\subsection{Crustal properties}

\begin{figure}[t!]
 \centering
 \includegraphics[width=1\linewidth, clip=true]{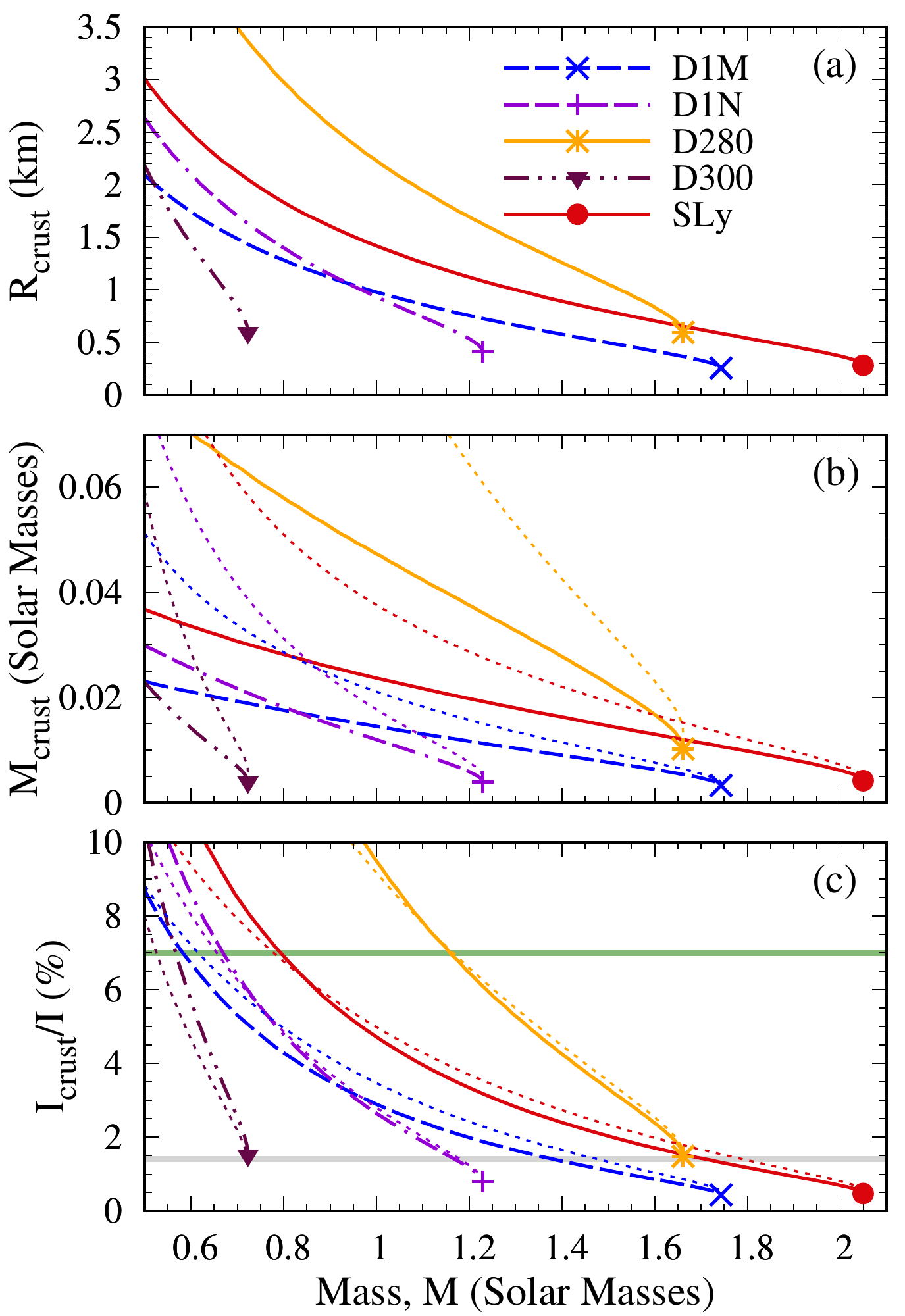}\\
   \caption{Panel (a): Crust thickness for the four stable Gogny functionals and the SLy EoS. Panel (b): Mass enclosed by the crust. Short-dashed lines correspond to the approximation of Eq.~(\ref{eq:crust_mass}). Panel (c): Percentage fraction of the star's moment of inertia contained in the crust. Short-dashed lines correspond to the approximation of Eq.~(\ref{eq:crust_momI}). Thick horizontal lines indicate the constraints of Refs.~\cite{Link1999} (bottom line) and~\cite{Andersson2012} (top line) to account for observed glitches in the Vela pulsar and other glitching sources. }
 \label{fig:crust}
\end{figure}

The solution of the TOV and moment of inertia equations, combined with the determination of the core-crust transition, allows us to separate the crust and the core within the neutron star \cite{Chamel2008}. One can, for instance, find the crust thickness, $R_\text{crust}$, which corresponds to the radial coordinate at which the crust-core transition takes place measured from the surface of the star. 
Similarly, the crust mass, $M_\text{crust}$, is the fraction of the star's mass enclosed by the crust. Finally, the crust moment of inertia, $I_\text{crust}$, is the fraction of moment of inertia within the star's crust. 
While the EoSs of the Gogny forces that we present are relatively soft and incompatible with the observations of the heaviest neutron stars~\cite{Demorest2010,Antoniadis2013}, one might expect the low-density physics around the core-crust transition to be well described by these functionals. 
We provide an overview of the crust properties for the considered Gogny parametrizations in Fig.~\ref{fig:crust}. In all the panels, we find that the SLy results lie within the Gogny D280 and D1M predictions. We take this as an indication of the fact that some Gogny forces indeed provide a relatively realistic description of the crust. Accordingly, in the following we concentrate on discussing the predictions from the D280 and D1M EoSs. 

Panel (a) of Fig.~\ref{fig:crust} summarises our results on the crust thickness. As expected, the thickness decreases with the mass of the star 
\cite{Chamel2008,Li2016}. For a canonical mass $M=1.4 M_\odot$, D1M predicts $R_\text{crust} \approx 0.6$ km whereas both D280 and SLy have larger 
crusts, $R_\text{crust} \approx 0.9-1.2$ km. In fact, D280 provides a significantly thicker neutron star crust than D1M for the whole mass region. Within the small number of forces that are available, it appears that models with a larger slope parameter $L \approx 45$ MeV, like D280 and SLy, produce larger crusts for a given mass. This is in principle in contrast to the 
systematics of Ref.~\cite{AJ697Xu2009}, although the results in that reference are quoted for $L > 60$~MeV.

The amount of mass contained in the crust is shown in panel (b) of Fig.~\ref{fig:crust}. We find that the crust mass decreases as the mass of the star increases 
\cite{Chamel2008,Li2016}. There is a large sensitivity to the EoS in the crust mass. D280, for instance, provides substantially larger crust 
masses than SLy and D1M in a wide range of masses. Moreover, the dependence in mass is steep for D280, whereas it is relatively flat 
for SLy and D1M. A canonical $1.4 M_\odot$ pulsar would have a crust mass $M_\text{crust} \approx 0.01 M_\odot$ ($\approx 0.03 M_\odot$) for D1M (D280), in between the SLy prediction of $M_\text{crust} \approx 0.02 M_\odot$.

Within a thin-crust approximation, the mass of the crust can be estimated by the expression \cite{Chamel2008}:
\begin{align}
M_\text{crust} \approx \frac{4 \pi R^4 P_t}{GM} \left[ 1 - 2 \chi \right] \, .
\label{eq:crust_mass}
\end{align}
This involves the pressure at the transition point, $P_t$, as well as properties computed at the surface of the star (total mass and radius). The results of this approximation (thin short-dashed lines) are compared to those obtained in the full TOV calculation in panel (b) of Fig.~\ref{fig:crust}. We note that the approximation overestimates the crust mass, particularly at low masses. 
In contrast, near the maximum mass configuration, the results of Eq.~(\ref{eq:crust_mass}) become closer to the exact ones. Above $M\approx1.4 M_\odot$, for instance, the approximation is good to within $\approx 0.01 M_\odot$. 
In fact, the mass and thickness of the crust can be entirely determined to excellent accuracy by the core EoS and the crust-core transition point as recently discussed 
in Ref.~\cite{Zdunik2017}.

We have explored the sensitivity of our results to a different treatment of the inner crust, by using the extended SLy EoS in this region instead of a polytropic parametrization. 
In general, for the analyzed forces we find moderate variations in the mass and radius of the crust with the inner crust treatment. 
For $M_\text{crust}$, we find a variation which is less than $2 \%$ for $M > M_\odot$. Similarly, $R_\text{crust}$ changes by about $10 \%$ for D1M, D1N, and D300, and by about $20 \%$ for D280, in the region where $M > M_\odot$. 
Ideally, the inner crust EoS should be computed with the same nuclear force used for the homogeneous matter of the core \cite{Than2011,Baldo:2013ska,Fortin:2016hny}, but this goes beyond the scope of this work.

Finally, we present in panel (c) of Fig.~\ref{fig:crust} the results for the crustal fraction of the moment of inertia. We find similar trends 
to those present in previous panels. $I_\text{crust}/I$ decreases with the pulsar mass. Up to about 
$M\approx 1.6 M_\odot$, D280 stands out above the other models. This is to be expected, as it predicted thicker 
and heavier crusts. Both D1M and SLy predict crustal fractions which are below $5 \%$ above $1 M_\odot$, 
whereas D280 only falls below this value above $1.3 M_\odot$. In fact, for a canonical pulsar with $M=1.4 M_\odot$, 
we find $I_\text{crust}/I\approx 4 \%$ for D280 and $ \approx 1 \%$ for D1M. 

An approximated formula for the crustal fraction of moment of inertia
is given by \cite{Link1999,AJ550LattimerPrakash2001,AJ697Xu2009}:
\begin{align}
\frac{I_\text{crust}}{I} \approx& \; \frac{28 \pi R^3 P_t}{3M} \, \frac{1 - 1.67 \chi -0.6 \chi^2}{\chi}  \nonumber \\
&\times \left[ 1 + \frac{2 P_t (1 + 5 \chi -14 \chi^2)}{m\rho_t \chi^2} \right]^{-1} \, .
\label{eq:crust_momI}
\end{align}
The results of this approximation are shown by the thin short-dashed lines in panel (c) of Fig.~\ref{fig:crust}. We find a very good agreement between the approximated formula and the full results above $1-1.2 M_\odot$, and the agreement improves as the mass of the pulsar increases. This is in keeping with the findings of Ref.~\cite{AJ697Xu2009}. 

To account for the sizes of observed glitches, the widely used pinning model requires that a certain amount of angular momentum is 
carried by the crust. This can be translated into constraints on the crustal fraction of the moment of inertia. 
Initial estimates suggested that $I_\text{crust}/I>1.4 \, \%$ to explain  Vela and other glitching sources 
\cite{Link1999}. We show this value as the bottom horizontal line in panel (c) of Fig.~\ref{fig:crust}. We note that this does not pose 
mass constraints on D280, which has a minimum value of $I_\text{crust}/I$ slightly above that limit. For D1M, in contrast, glitching sources that satisfy this constraint should have $M<1.4M_\odot$. 
More recently, a more stringent constraint has 
been obtained by accounting for the entrainment of neutrons in the crust \cite{Andersson2012}. With entrained neutrons, a 
larger crustal fraction of moment of inertia, $I_\text{crust}/I>7 \, \%$ (top horizontal line in panel (c)), is needed 
to explain glitches. For D280, this represents a mass constraint below $M<1.1 M_\odot$. In contrast, D1M would need 
significantly lower masses, $M<0.6 M_\odot$, to account for glitching phenomena. Of course, a more realistic account 
of nuclear structure and superfluidity in the crust will modify the estimates. In particular, Gogny forces, which 
can naturally account for superfluidity, would be helpful in the modeling of the microphysics of neutron star crusts. 

\section{Summary and outlook}
\label{sec:summ}

In this paper, we attempt to link the microphysical predictions associated to the isospin dependence of the Gogny interaction to the observational properties of the neutron star core-crust transition. 
On the one hand, we investigate the influence of the symmetry energy on 
the core-crust transition using different Gogny forces. On the other hand, we study the stellar masses and radii predicted by the Gogny forces, paying special attention to properties related with the crust, such as its thickness, mass and fraction of the moment of inertia, which can have observational consequences. These properties are directly related to the core-crust transition, which can be computed in the thermodynamical method.

We first analyze the Taylor expansion of the energy per particle of asymmetric nuclear matter in even powers of the isospin asymmetry $\delta$. The lowest order is the contribution in 
symmetric nuclear matter and the next term, quadratic in $\delta$, corresponds to the usual symmetry energy coefficient. Higher-order
 terms in the Taylor expansion provide additional corrections that account for the departure of the energy from a quadratic law in $\delta$. 
The second-order symmetry energy coefficient in the analyzed Gogny interactions shows a well-known isospin instability at large values of the density, above $0.4 - 0.5$~fm$^{-3}$.
The fourth- and sixth-order symmetry energy coefficients contain contributions from the kinetic and
exchange terms exclusively. The results indicate that Gogny parametrizations fall into two different groups according to the density behavior of these coefficients above saturation.
In the first group (D1S, D1M, D1N, and D250), the fourth- and sixth-order coefficients reach a maximum and then decrease with growing density. In the second group (D1, D260, D280, and D300), these coefficients are always increasing functions of density in the range analyzed. The different behavior
of the two groups can be traced back to the density dependence of the exchange terms, which  
add to the kinetic part of the fourth- and sixth-order coefficients. At saturation density, the fourth- and sixth-order symmetry energy coefficients
are relatively small. This supports the accuracy 
of the Taylor expansion at second order in calculations of the energy in asymmetric nuclear matter around this density. 

The symmetry energy is often evaluated through the so-called parabolic approximation, as the difference between the energy per particle in pure neutron matter and in symmetric matter. 
We find that around saturation the difference between the PA estimate $E_{\mathrm{sym}}^{PA} (\rho)$ and the $E_{\mathrm{sym},2} (\rho)$ coefficient is largely accounted by the sum of the fourth- and sixth-order contributions.
Another important quantity in studies of the symmetry energy is the slope parameter $L$, which is commonly used to characterize the density dependence of the symmetry energy near saturation.
We find that large discrepancies of several MeV can arise between the $L$ value calculated with $E_{\mathrm{sym},2} (\rho)$ or with $E_{\mathrm{sym}}^{PA} (\rho)$, particularly for group~2 forces. Again, adding the fourth- and sixth-order contributions accounts for most of these differences.

To study the core-crust transition in neutron stars, one needs to consider $\beta$-stable stellar matter first. We take into account neutrons, protons and electrons in chemical equilibrium. By solving the equations with the exact EoS and with the Taylor expansion of Eq.~(\ref{EoS}) at increasing orders in $\delta$, we are able to analyze the convergence of the solutions with the expansion. The corresponding isospin asymmetry for $\beta$-stability
is always close to $\delta \approx 1$, in accord with the relatively soft symmetry energies associated to Gogny forces. The agreement between the $\beta$-equilibrium asymmetries obtained using the exact EoS and the truncated Taylor expansion improves order by order.
However, the convergence of this expansion is rather slow, in particular for forces with larger slope parameters~$L$. 

The core-crust transition density is estimated using the thermodynamical method. 
The change of sign of the potential $V_{\mathrm{ther}} (\rho)$ determines the onset of instabilities. In general, adding more terms to the Taylor expansion of the EoS brings the transition density closer to the value of the exact EoS.
However, there can be still significant differences even when the Taylor expansion is pushed to sixth order. This points out that the convergence for the transition properties is slow. As noted in earlier literature, at least for Skyrme forces and RMF parameter sets,
the core-crust transition density is anticorrelated with the slope parameter $L$ of the models. Our calculations confirm this trend for Gogny forces also.
Although we have a reduced number of forces, if we take into account their slope parameters, the predictions are consistent with the expected window of values provided by the Skyrme and RMF models. In contrast to the transition density, the transition pressure analyzed with Gogny forces is not seen to correlate with~$L$.

Next, we have studied several neutron star properties using Gogny interactions. We find that only the 
D1M, D1N, D280, and D300 forces provide numerically stable solutions of the TOV equations. The maximum mass configurations for D1M and D280 occur at  $M=1.74 M_\odot$ and $1.66 M_\odot$, respectively, clearly below the observational limit of $2 M_\odot$. In contrast, D1N and D300 predict neutron stars with maximum masses below the canonical value $1.4 M_\odot$, as well as unrealistically small radii. The central densities of both maximum and canonical mass neutron stars computed with D1M and D280 are rather large. This is consistent 
with the soft neutron matter EoSs of these interactions. Another quantity of interest 
is the moment of inertia of the star, which has a maximum value of $I_\text{max} \approx 1.3-1.4 \times 10^{45}$ g cm$^2$ in the D1M and D280 forces.

The solution of the TOV and moment of inertia equations, together with the core-crust transition density, allows one to 
predict the crust thickness and to separate the mass and the moment of inertia into crust and core contributions. 
Although some of the bulk stellar properties predicted by the Gogny forces are incompatible with 
 observations, the physics around the core-crust transition seems to be rather well described by D280 and, in particular, by D1M, which gives results commensurate with previous literature, and similar to those obtained with the SLy EoS. 
Finally, let us point out that this is not completely surprising. Gogny forces are fitted
to nuclear properties at relatively low densities and close to isospin symmetric conditions. High-density neutron-rich systems are normally beyond the fit of these forces. However, the relatively low-density physics of the core-crust transition can be well described as long as the near-saturation isospin dependence is realistic. It appears that D1M performs relatively well in this context.
        
From the present analysis of neutron stars with Gogny forces, we see that there is room for improvements.
On the one hand, the so-called dynamical method has often been used to compute the core-crust transition with Skyrme forces \cite{NPA175BAYM1971,NPA584Pethick1995,AJ697Xu2009,NPA789Ducoin2007,PRC83Ducoin2011}.
With a proper extension, this method could be generalized to the case of Gogny interactions. Moreover, quantum-mechanical predictions of random phase approximation instabilities in infinite matter are now available, and their extension to isospin asymmetric matter should be an informative step forward \cite{DePace2016}. 
On the other hand, it would be desirable to construct new Gogny parametrizations which are able to reproduce simultaneously finite nuclei and the most recent constraints from neutron star observations. There is no reason why new parametrizations could not achieve a similar quality to the SLy or BSk families of Skyrme forces.

\acknowledgments
C.G., M.C., and X.V. acknowledge support from Grant FIS2014-54672-P from MINECO and FEDER,
Grant 2014SGR-401 from Generalitat de Catalunya,
and Project MDM-2014-0369 of ICCUB (Unidad de Excelencia Mar\'{\i}a de Maeztu) from MINECO.
C.G. also acknowledges Grant BES-2015-074210 from MINECO.
The work of A.R. was supported by STFC through Grants ST/I005528/1, ST/J000051/1, ST/L005743/1  and ST/L005816/1. 
Partial support came from ``NewCompStar", COST Action MP1304.


\appendix

\section{Total baryon energy and symmetry energy up to sixth order for Gogny forces}
\label{appendix1}

This appendix contains the expression of the baryon energy per particle $E_b (\rho, \delta)$ in asymmetric nuclear matter with Gogny interactions. We also give the expressions for
the symmetry energy coefficients entering the Taylor expansion of $E_b (\rho, \delta)$ [see Eq.~(\ref{EoS})] through sixth order in the isospin asymmetry $\delta$, and the respective slope parameters defined in Eq.~(\ref{eq:L2k}). 

The total energy per particle in the Hartree--Fock approximation with the Gogny two-body effective interaction given in Eq.~(\ref{VGogny}) becomes the sum of four different contributions, namely, a kinetic contribution, a zero-range contribution, and the direct and exchange contributions:
\begin{eqnarray}
 E_b (\rho, \delta)&=&  E_b^{\mathrm{kin}} (\rho, \delta)+ E_b^{\mathrm{zr}} (\rho, \delta) \nonumber \\ 
 && \mbox{} + E_b^{\mathrm{dir}} (\rho, \delta) + E_b^{\mathrm{exch}} (\rho, \delta) \label{eq:eb.terms} \, ,
\end{eqnarray}
which read as
\begin{widetext}
\begin{eqnarray}
 E_b^{\mathrm{kin}} (\rho, \delta)&=& \frac{ 3 \hbar^2}{20m} \left(\frac{3 \pi^2}{2}\right)^{2/3} \rho^{2/3}
 \left[ (1+\delta)^{5/3} + (1-\delta)^{5/3} \right] \label{eq:eb.kin}
\\
E_b^{\mathrm{zr}} (\rho, \delta)&=&  \frac{1}{8} t_3 \rho^{\alpha+1} \left[ 3-(2x_3+1)\delta^2 \right] \label{eq:eb.zr}
\\
E_b^{\mathrm{dir}} (\rho, \delta)&=&  \frac{1}{2} \sum_{i=1,2} \mu_i^3 \pi^{3/2} \rho  \left[ {\cal A}_i 
+{\cal B}_i \delta^2 \right] \label{eq:eb.dir}
\\
E_b^{\mathrm{exch}} (\rho, \delta)&= & -\sum_{\mathrm{i}=1,2}\frac{1}{2  k_F^3 \mu_i^3} 
\Big\{ {\cal C}_i \left[ {\mathsf  e} (k_{Fn} \mu_i ) + {\mathsf  e} (k_{Fp} \mu_i ) \right]
-  {\cal D}_i  \bar {\mathsf  e}( k_{Fn} \mu_i,k_{Fp} \mu_i )  \Big\}, \label{eq:eb.exch}
\end{eqnarray}
 with 
 \begin{equation}
{\mathsf e}(\eta) = \frac{\sqrt{\pi}}{2} \eta^3 \mathrm{erf}(\eta) 
+ \left(\frac{\eta^2}{2} - 1 \right) e^{-\eta^2} - \frac{3 \eta^2}{2} +  1 \, ,
\end{equation}
and 
 \begin{equation}
\bar {\mathsf  e}(\eta_1,\eta_2)= \sum_{s=\pm 1} s 
\left[ 
\frac{ \sqrt{\pi}}{2} (\eta_1 + s \eta_2 ) \left( \eta_1^2 + \eta_2^2 - s \eta_1 \eta_2 \right) 
\mathrm{erf} \left( \frac{\eta_1 + s \eta_2 }{2}  \right) 
+ \left( \eta_1^2 + \eta_2^2  - s \eta_1 \eta_2  -2 \right) e^{ - \frac{1}{4} (\eta_1 + s \eta_2)^2 } 
\right] ,
\end{equation}
where $\displaystyle \mathrm{erf}(x) =  \frac{2}{\sqrt{\pi}} \int_0^x  e^{-t^2} dt $ is the error function.
The function $\bar {\mathsf  e}(\eta_1,\eta_2)$ is a symmetric function of its arguments, satisfying $\mathsf{\bar e}(\eta,\eta)= 2 \mathsf{e}(\eta)$ and $\mathsf{\bar e}(\eta,0) = 0$.

The term in Eq.~(\ref{eq:eb.kin}) is the sum of the neutron and proton kinetic energy contributions, whereas the zero-range term in Eq.~(\ref{eq:eb.zr}) is the contribution of the contact interaction. 
Both can be expressed in terms of the total baryon density $\rho$ and the isospin asymmetry in the system $\delta = (\rho_n-\rho_p)/(\rho_n + \rho_p)$.
The direct term in Eq.~(\ref{eq:eb.dir}) and the exchange term in Eq.~(\ref{eq:eb.exch}) are the contributions to the energy from the finite range part of the Gogny interaction.
The direct term $E_b^{\mathrm{dir}} (\rho, \delta)$ is easily expressed in terms of the density $\rho$ and of $\delta^2$. The exchange term $E_b^{\mathrm{exch}} (\rho, \delta)$, in contrast, is a function of the neutron and proton Fermi momenta:
$k_{Fn} = k_F (1+\delta)^{1/3}$ and  $k_{Fp} = k_F (1-\delta)^{1/3}$, respectively. The Fermi momentum 
of isospin symmetric matter is given by $k_F = (3 \pi^2 \rho/2 )^{1/3}$.
The following combinations have been used in order to present the finite range terms:
\begin{eqnarray}
{\cal A}_i &=& \frac{1}{4} \left( 4 W_i + 2 B_i - 2H_i -M_i \right) \label{Ai}
\\
{\cal B}_i&=&  -\frac{1}{4}\left( 2 H_i + M_i \right)\label{Bi}
\\
{\cal C}_i&=& \frac{1}{\sqrt{\pi}} \left( W_i + 2 B_i - H_i -2 M_i\right)\label{Ci}
\\
{\cal D}_i&=& \frac{1}{\sqrt{\pi}} \left( H_i + 2 M_i \right).\label{Di}
\end{eqnarray}
The constants $ {\cal A}_i $ and ${\cal B}_i$ define, respectively, the isoscalar and isovector 
part of the direct term. For the exchange terms, the matrix elements ${\cal C}_i $ relate to the interaction 
between particles with the same isospin (neutron-neutron and proton-proton interactions), whereas the matrix elements ${\cal D}_i$ take care of interactions between particles with different isospin (neutron-proton interactions). 

From the energy per baryon we can obtain analytical expressions for the symmetry energy coefficients, which up to sixth order are:
\begin{eqnarray}
E_{\mathrm{sym}, 2} (\rho) = 
\left. \frac{1}{2!} \frac{\partial^{2} E_b(\rho, \delta)}{\partial \delta^{2}}\right|_{\delta=0} &=&
\frac{\hbar^2}{6m} \left(  \frac{3 \pi^2}{2}\right)^{2/3} \rho^{2/3}  - 
\frac{1}{8} t_3 \rho^{\alpha+1} (2x_3 +1) + \frac{1}{2} \sum_{i=1,2} \mu_i^3 \pi^{3/2}  {\cal B}_i  \rho  \nonumber
\\
&& \mbox{} +\frac{1}{6}\sum_{i=1,2}  \left[-{\cal C}_i  G_1 ( k_F \mu_i)+ {\cal D}_i G_2 ( k_F \mu_i)  \right] , \label{eq:esym2}
\\
E_{\mathrm{sym}, 4} (\rho) =
\left. \frac{1}{4!} \frac{\partial^{4} E_b(\rho, \delta)}{\partial \delta^{4}}\right|_{\delta=0} &=&
\frac{\hbar^2}{162m} \left(  \frac{3 \pi^2}{2}\right)^{2/3} \rho^{2/3} +
\frac{1}{324} \sum_{i=1,2}
\left[ {\cal C}_i  G_3 ( k_F \mu_i)  + {\cal D}_i G_4 ( k_F \mu_i) \right] \, ,\label{eq:esym4}
\\
E_{\mathrm{sym}, 6} (\rho) =
\left. \frac{1}{6!} \frac{\partial^{6} E_b(\rho, \delta)}{\partial \delta^{6}}\right|_{\delta=0} &=&
\frac{7\hbar^2}{4374m} \left(  \frac{3 \pi^2}{2}\right)^{2/3} \rho^{2/3} + 
\frac{1}{43740} \sum_{i=1,2}
 \left[ {\cal C}_i  G_5(k_F \mu_i )  - {\cal D}_i  G_6( k_F \mu_i) \right] , \label{eq:esym6}
\end{eqnarray}
with
\begin{eqnarray}
 G_1 (\eta)&=& \frac{1}{\eta} -\left( \eta + \frac{1}{\eta} \right) e^{-\eta^2} \label{G1}
\\
 G_2 (\eta)&=& \frac{1}{\eta} -\bigg( \eta + \frac{e^{-\eta^2}}{\eta} \bigg) \label{G2}
\\
 G_3 (\eta)&=&  -\frac{14}{\eta}  + e^{-\eta^2} \left( \frac{14}{\eta} + 14 \eta + 7 \eta^3 + 2\eta^5 \right) \label{G3}
\\
 G_4 (\eta)&=&   \frac{14}{\eta} - 8 \eta + \eta^3 - 2 e^{-\eta^2} \left( \frac{7}{\eta}+3\eta\right)\label{G4}
\\
 G_5 (\eta)&=&  -\frac{910}{\eta} + e^{-\eta^2}\left( \frac{910}{\eta} +   910\eta  + 455 \eta^3 + 147 \eta^5 
 + 32\eta^7 + 4 \eta^9 \right) \label{G5}
\\
 G_6 (\eta)&=&  -\frac{910}{\eta} + 460 \eta - 65 \eta^3+ 3 \eta^5 +e^{-\eta^2} \left( \frac{910}{\eta} 
 + 450 \eta + 60\eta^3    \right). \label{G6}
\end{eqnarray}
The corresponding slope parameters $L \equiv L_2$, $L_4$, and $L_6$ at saturation density $\rho_0$ are given by
 \begin{eqnarray}
 L = 3 \rho_0 \left. \frac{\partial E_{\mathrm{sym},2} (\rho)}{\partial \rho} \right|_{\rho_0}&=& \frac{\hbar^2}{3m} \left(\frac{3 \pi^2}{2}\right)^{2/3} \rho_0^{2/3} -
  \frac{3(\alpha + 1)}{8} t_3 \rho_0^{\alpha+1} (2 x_3 +1) + \frac{3}{2} \sum_{i=1,2} \mu_i^3 \pi^{3/2} \mathcal{B}_i \rho_0 \nonumber \\
  && \mbox{} + \frac{1}{6} \sum_{i=1,2} \mu_i k_{F0} \left[ - \mathcal{C}_i G_1'(\mu_i k_{F0}) + \mathcal{D}_i G_2'(\mu_i k_{F0}) \right] ,
  \\
 L_4 = 3 \rho_0 \left. \frac{\partial E_{\mathrm{sym},4} (\rho)}{\partial \rho} \right|_{\rho_0} &=& \frac{\hbar^2}{81m} \left(\frac{3 \pi^2}{2}\right)^{2/3} \rho_0^{2/3}
+ \frac{1}{324} \sum_{i=1,2} \mu_i k_{F0} \left[ \mathcal{C}_i G_3'(\mu_i k_{F0}) + \mathcal{D}_i G_4'(\mu_i k_{F0}) \right] ,
\\
 L_6 = 3 \rho_0 \left. \frac{\partial E_{\mathrm{sym},6} (\rho)}{\partial \rho} \right|_{\rho_0}&=& \frac{7\hbar^2}{2187m} \left(\frac{3 \pi^2}{2}\right)^{2/3} \rho_0^{2/3} 
+ \frac{1}{43740} \sum_{i=1,2} \mu_i k_{F0} \left[ \mathcal{C}_i G_5'(\mu_i k_{F0}) - \mathcal{D}_i G_6'(\mu_i k_{F0}) \right] ,
 \end{eqnarray}
where $k_{F0} = (3 \pi^2 \rho_0/2)^{1/3}$ is the Fermi momentum at saturation and the derivatives of the $G_n (\eta)$ functions are
\begin{eqnarray}
 G_1' (\eta) &=& -\frac{1}{\eta^2} +e^{-\eta^2} \left( \frac{1}{\eta^2} + 1 + 2 \eta^2\right) \label{G1'}
 \\
 G_2' (\eta) &=&  -\frac{1}{\eta^2} + e^{-\eta^2} \left( \frac{1}{\eta^2} +2 \right) - 1 
 \\
  G_3' (\eta) &=&  \frac{14}{\eta^2} - e^{-\eta^2}\left(\frac{14}{\eta^2} + 14 + 7 \eta^2 + 4 \eta^4 + 4 \eta^6 \right)
  \\
 G_4' (\eta) &=&  -\frac{14}{\eta^2} - 8 + 3 \eta^2 + e^{-\eta^2} \left(\frac{14}{\eta^2} + 22 + 12 \eta^2 \right)
 \\
 G_5' (\eta) &=&  \frac{910}{ \eta^2}  - e^{-\eta^2} \left( \frac{910}{ \eta^2} +910 + 455 \eta^2 +175 \eta^4 + 70 \eta^6 + 28 \eta^8 +8 \eta^{10} \right)
 \\
 G_6' (\eta) &=&   \frac{910}{\eta^2} + 460 -195 \eta^2 + 15 \eta^4 - e^{-\eta^2} \left(\frac{910}{\eta^2} + 1370+720 \eta^2 + 120 \eta^4 \right) . \label{G6'}
\end{eqnarray}

\section{Chemical potentials and pressure in isospin asymmetric matter}
\label{appendix_p}
The neutron and proton chemical potentials in asymmetric nuclear matter are the derivatives of the baryon energy
density $\mathcal{H}_b$ with respect to the neutron or proton densities, respectively, cf.
Eq.~(\ref{eq:potentials}). With $\tau = +1$ for neutrons and $\tau = -1$ for protons, the nucleon 
chemical potentials for the Gogny interaction are given by
\begin{eqnarray}
 \mu_\tau &=& \frac{\hbar^2}{2m} \left( 3 \pi^2 \right)^{2/3} \rho_\tau^{2/3} + \frac{t_3}{8} \rho^{\alpha + 1} 
 \left[ 3 \left( \alpha + 2\right) - 2 \tau \left( 2 x_3 +1\right) \delta - \alpha \left( 2 x_3 + 1\right) 
 \delta^2\right] \nonumber
 \\
 && + \sum_{i=1,2} \mu_i^3 \pi^{3/2} \rho \left( {\cal A}_i + \tau {\cal B}_i \delta \right) 
 - \sum_{i=1,2} \left[ {\cal C}_i\,  \bar {\mathsf w}( k_F^\tau \mu_i , k_F^\tau \mu_i  ) - 
 {\cal D}_i\, \bar {\mathsf  w} (k_F^\tau \mu_i , k_F^{-\tau} \mu_i )\right],
\label{mu-tau}
\end{eqnarray}
where $\bar {\mathsf w} \left( \eta_1 , \eta_2\right)$ is the dimensionless function 
\begin{equation}
\bar {\mathsf w} \left( \eta_1 , \eta_2\right) =
\sum_{s=\pm 1} s \left[ \frac{\sqrt{\pi}}{2} \mathrm{erf}\left( \frac{\eta_1 + s \eta_2}{2}\right) 
+ \frac{1}{\eta_1} e^{- \frac{1}{4} ( \eta_1 + s \eta_2 )^2} \right] \, .
\end{equation}

The baryon pressure in isospin asymmetric matter can be obtained from the derivative of $E_b (\rho, \delta)$ with respect to the baryon density, Eq.~(\ref{eq:Pb}). It may also be computed from the chemical potentials and the baryon energy density following Eq.~(\ref{PbPe}). Either way, one finds:
\begin{eqnarray}
P_b(\rho,\delta) &=&
\frac{\hbar^2}{10m} \left(\frac{3 \pi^2}{2}\right)^{2/3} \rho^{5/3} \left[ (1+\delta)^{5/3} + (1-\delta)^{5/3} 
\right]
+ \frac{(\alpha+1)}{8} t_3 \rho^{\alpha+2} \left[ 3 - (2x_3+1) \delta^2 \right] \nonumber \\
& & + \frac{\rho^2}{2} \sum_{i=1,2} \pi^{3/2} \mu_i^3 \left( {\cal A}_i + {\cal B}_i \delta^2 \right) \nonumber \\
& & - \frac{\rho}{2}  \sum_{i=1,2} \left\{ 
{\cal C}_i
\left[ (1+\delta) \mathsf{p} ( k_{Fn} \mu_i ) 
     + (1-\delta) \mathsf{p} ( k_{Fp} \mu_i ) \right]   
- {\cal D}_i \mathsf{ \bar p} ( k_{Fn} \mu_i,k_{Fp}\mu_i )  
     \right\} \, .
     \label{eq:pressure_bars}
\end{eqnarray}
The function $\mathsf{p}(\eta)$ contains the density dependence of the pressure in both symmetric and neutron 
matter \cite{PRC90SellahewaArnauRios2014}:
\begin{align}
\mathsf{p}\left(\eta \right) &= -\frac{1}{\eta^3} + \frac{1}{2 \eta} + \left( \frac{1}{\eta^3} + 
\frac{1}{2 \eta} \right) e^{- \eta^2} \, .
\end{align}
In asymmetric matter, the double integral on the exchange terms leads to the appearance of a term that 
depends on the two Fermi momenta: 
\begin{align}
 \mathsf{\bar p} \left( \eta_1, \eta_2 \right)  = &
\frac{2}{\eta_1^3 + \eta_2^3}
\sum_{s=\pm 1}  (\eta_1 \eta_2 +2s ) e^{- \frac{1}{4} \left(\eta_1 + s\eta_2\right)^2 } .
 \label{eq:p_function}
 \end{align}
This term is a symmetric function of its arguments, which fulfils
$\mathsf{\bar p}(\eta,\eta)= 2 \mathsf{p}(\eta)$ and $\mathsf{\bar p}(\eta,0) = 0$.

\section{Thermodynamical Potential}
\label{appendix_thermal}
The stability condition for the thermodynamical potential $V_\mathrm{ther} (\rho)$ discussed in Sec.~\ref{sec:thermodynamicalmethod} requires the calculation of
the first and second derivatives of the Gogny energy per baryon $E_b (\rho, \delta)$ with respect to density $\rho$ and isospin asymmetry $\delta$. 
In this appendix we provide the corresponding expressions obtained with the exact EoS and with the Taylor expansion of the EoS up to order $\delta^6$.

\subsection{Derivatives for $V_\mathrm{ther} (\rho)$ using the exact expression of the EoS}
Here, we collect the derivatives of $E_b (\rho, \delta)$ involved in the stability condition V$_\mathrm{ther} (\rho) >0$ in Eq.~(\ref{Vthermal}). 
The derivative $\partial E_b (\rho, \delta)/\partial \rho$ is immediately obtained from the expression for the pressure $P_b(\rho,\delta)$ we have given in Eq.~(\ref{eq:pressure_bars}), taking into account that $\partial E_b (\rho, \delta)/\partial \rho = P_b(\rho,\delta)/\rho^2$. The other derivatives that appear in Eq.~(\ref{Vthermal}) are:
%
\begin{eqnarray}
\frac{\partial^2 E_b (\rho, \delta)}{\partial \rho^2}&=& - \frac{\hbar^2}{30m} \left(\frac{3 \pi^2}{2} 
\right)^{2/3} \rho^{-4/3}\left[ (1+\delta)^{5/3} + (1-\delta)^{5/3} \right]
+ \frac{(\alpha + 1) \alpha}{8} t_3 \rho^{\alpha-1} \left[ 3 - \left( 2 x_3 + 1 \right) \delta^2 \right] 
\nonumber
\\
&& +\sum_{i=1,2}\frac{1}{6 \rho^2 k_F^{3} \mu_i^3}  \Bigg\{  {\cal C}_i \left[ 
\vphantom{\frac{1}{2}}2 \left( -6 + 
k_{Fn}^2\mu_i^2 + k_{Fp}^2\mu_i^2\right) +  e^{- k_{Fn}^2\mu_i^2} \left( 6  + 4  k_{Fn}^2\mu_i^2 + 
k_{Fn}^4\mu_i^4\right)  \right.  \nonumber
\\
&& \left.
 + e^{- k_{Fp}^2\mu_i^2} \left( 6  + 4  k_{Fp}^2\mu_i^2 + k_{Fp}^4\mu_i^4\right) \vphantom{\frac{1}{2}} \right]
 + {\cal D}_i e^{-\frac{1}{4} \left( k_{Fn}^2 + k_{Fp}^2\right)\mu_i^2} \nonumber
\\
&& \times \left[  \left( -12  k_{Fn} k_{Fp}\mu_i^2- k_{Fn}^3 k_{Fp}\mu_i^4 - k_{Fn} k_{Fp}^3\mu_i^4 \right) 
\mathrm{cosh} \left[ \frac{k_{Fn} k_{Fp}\mu_i^2}{2}  \right] \right.  \nonumber
\\
&& + 2 \left( 12 + k_{Fn}^2\mu_i^2  +  k_{Fp}^2\mu_i^2 + k_{Fn}^2 k_{Fp}^2\mu_i^4\right) 
\left.  \mathrm{sinh} \left[ \frac{k_{Fn} k_{Fp}\mu_i^2}{2} \right] \right] \Bigg\} ,
\end{eqnarray}
\begin{eqnarray}
\frac{\partial^2 E_b (\rho, 
\delta)}{\partial \rho \partial \delta} &=&  \frac{\hbar^2}{6m} \left(\frac{3 \pi^2}{2} \right)^{2/3} 
\rho^{-1/3}\left[ (1+\delta)^{2/3} - (1-\delta)^{2/3} \right]
- \frac{(\alpha +1)}{4} t_3 \rho^{\alpha} (2x_3+1) \delta  +  \sum_{i=1,2} \mu_i^3 \pi^{3/2} {\cal B}_i\delta \nonumber
\\
&& -\sum_{i=1,2}\frac{1}{6 \rho} \left\{{\cal C}_i \left[ \frac{-1 + e^{-k_{Fp}^2 \mu_i^2} 
\left( 1+ k_{Fp}^2 \mu_i^2\right)}{k_{Fp} \mu_i} - \frac{-1 + e^{-k_{Fn}^2 \mu_i^2} 
\left( 1+ k_{Fn}^2 \mu_i^2\right)}{k_{Fn} \mu_i} \right]\nonumber \right. \nonumber
\\
&& -{\cal D}_i e^{-\frac{1}{4} \left(k_{Fn}^2+ k_{Fp}^2\right)\mu_i^2} 
\left[ \left(k_{Fn} \mu_i - k_{Fp} \mu_i \right)
\cosh \left[ \frac{k_{Fn} k_{Fp}\mu_i^2}{2} \right] \right. \nonumber
\\
&& \left. - \frac{2}{k_{Fn} k_{Fp} \mu_i^2}  \left( k_{Fn} \mu_i - k_{Fp} \mu_i + \delta k_F^3 \mu_i^3\right)
\sinh \left[  \frac{k_{Fn} k_{Fp}\mu_i^2}{2} \right] \right] \Bigg\} ,
\end{eqnarray}
\begin{eqnarray}
\frac{\partial^2 E_b (\rho, \delta)}{ \partial \delta^2} &=& 
\frac{\hbar^2}{6m} \left(\frac{3 \pi^2}{2} \right)^{2/3} \rho^{2/3}\left[ (1+\delta)^{-1/3} + (1-\delta)^{-1/3} \right]
-  \frac{t_3}{4} \rho^{\alpha+1} (2x_3+1) + \frac{1}{4}\sum_{i=1,2} \mu_i^3 \pi^{3/2}  {\cal B}_i \rho \nonumber
\\
&& -\frac{1}{6} \sum_{i=1,2} \left\{ {\cal C}_i \left [  \frac{1 - e^{- k_{Fp}^2 \mu_i^2}
\left( 1 + k_{Fp}^2 \mu_i^2 \right)}{(1-\delta) k_{Fp} \mu_i} + 
\frac{1 - e^{- k_{Fn}^2 \mu_i^2} 
\left( 1 + k_{Fn}^2 \mu_i^2 \right)}{(1+\delta) k_{Fn} \mu_i} \right] \right.\nonumber
\\
&& + {\cal D}_i  e^{-\frac{1}{4} \left( k_{Fp}^2 +k_{Fn}^2 \right) \mu_i^2}
 \left[ \left(  k_{Fn} \mu_i\left(1-\delta \right)^{-1} +  k_{Fp} \mu_i\left(1+\delta \right)^{-1}\right)  
\cosh \left[\frac{k_{Fn} k_{Fp}\mu_i^2}{2}\right] \right. \nonumber
\\
&&  - \frac{2}{\left( 1- \delta^2 \right) k_{Fn} k_{Fp} \mu_i^2} \left( k_{Fn} \mu_i + k_{Fp} \mu_i 
- k_F^3 \mu_i^3  \right. \nonumber
\\
&&\left.   \left. +\delta \left(  k_{Fn} \mu_i - k_{Fp} \mu_i   
+ \delta k_F^3 \mu_i^3 \right) \right)\sinh \left[\frac{k_{Fn} k_{Fp}\mu_i^2}{2}\right] \right] \Bigg\} .
\end{eqnarray}

\subsection{Derivatives for $V_\mathrm{ther} (\rho)$ using the Taylor expansion of the EoS}
If one replaces the EoS of asymmetric matter $ E_b (\rho, \delta)$ with its Taylor expansion in powers of the isospin asymmetry~$\delta$, the stability condition $V_\mathrm{ther} (\rho) >0$ takes the form shown in Eq.~(\ref{eq:Vtherapprox}). 
Expressing Eq.~(\ref{eq:Vtherapprox}) to sixth order in $\delta$ gives the result
\begin{eqnarray}
V_{\mathrm{ther}} (\rho) &=& \rho^2 \frac{\partial^2 E_b(\rho, \delta=0)}{\partial \rho^2} + 2  \rho \frac{\partial 
E_b(\rho, \delta=0)}{\partial \rho} 
+ \delta^2 \left( \rho^2 \frac{\partial^2 E_{\mathrm{sym}, 2}(\rho)}{\partial \rho^2} 
+ 2  \rho \frac{\partial E_{\mathrm{sym}, 2}(\rho)}{\partial \rho}\right) \nonumber
\\
&&+ \delta^4 \left( \rho^2 \frac{\partial^2 E_{\mathrm{sym}, 4}(\rho)}{\partial \rho^2}
+ 2  \rho \frac{\partial E_{\mathrm{sym}, 4}(\rho)}{\partial \rho}\right) 
+ \delta^6 \left( \rho^2 \frac{\partial^2 E_{\mathrm{sym}, 6}(\rho)}{\partial \rho^2} 
+ 2  \rho \frac{\partial E_{\mathrm{sym}, 6}(\rho)}{\partial \rho}\right) \nonumber
\\
&&  -\frac{2 \rho^2 \delta^2}{E_{\mathrm{sym}, 2}(\rho) + 6 E_{\mathrm{sym}, 4} (\rho)
\delta^2 + 15 E_{\mathrm{sym}, 6}(\rho) \delta^4 } \nonumber
\\
&&\times \left( \frac{\partial E_{\mathrm{sym}, 2}(\rho)}{\partial \rho}
+ 2 \delta^2 \frac{\partial E_{\mathrm{sym}, 4}(\rho)}{\partial \rho} +
3 \delta^4 \frac{\partial E_{\mathrm{sym}, 6}(\rho)}{\partial \rho} \right)^2 >0.
\label{vther-app}
\end{eqnarray}
For the Gogny interaction, the density derivatives of the energy per baryon in symmetric nuclear matter $E_b(\rho, \delta=0)$ that are needed for evaluating (\ref{vther-app}) are given by
\begin{eqnarray}
\frac{\partial E_b(\rho, \delta=0)}{\partial \rho} &=& \frac{\hbar^2}{5m} \left(  \frac{3 \pi^2}{2}\right)^{2/3} \rho^{-1/3} 
+ \frac{3(\alpha +1)}{8} t_3 \rho^{\alpha}   +
 \frac{1}{2} \sum_{i=1,2} \mu_i^3 \pi^{3/2}  {\cal A}_i \nonumber
\\
 &&-\sum_{i=1,2}\frac{1}{2 \rho k_F^3 \mu_i^3}  \left( {\cal C}_i - {\cal D}_i \right)
 \left[ -2 + k_F^2 \mu_i^2 + e^{-k_F^2 \mu_i^2} \left(2 + k_F^2 \mu_i^2 \right) \right] ,
\\
\frac{\partial^2 E_b(\rho, \delta=0)}{\partial \rho^2} &=& -\frac{\hbar^2}{15m} \left(  \frac{3 \pi^2}{2}\right)^{2/3} \rho^{-4/3} 
+ \frac{3(\alpha +1) \alpha}{8} t_3 \rho^{\alpha-1}\nonumber
\\  
&& -\sum_{i=1,2}\frac{1}{3 \rho^2 k_F^3 \mu_i^3}  \left( {\cal C}_i - {\cal D}_i \right)
\left[ 6 -2 k_F^2 \mu_i^2 - e^{-k_F^2 \mu_i^2} \left(6 +4 k_F^2 \mu_i^2 +k_F^4 \mu_i^4\right) \right] .
\end{eqnarray}
The first and second derivatives with respect to density of the symmetry energy coefficients $E_{\mathrm{sym}, 2}(\rho)$, $E_{\mathrm{sym}, 4}(\rho)$, 
and $E_{\mathrm{sym}, 6}(\rho)$ for the inequality (\ref{vther-app}) can be readily computed from Eqs.~(\ref{eq:esym2})--(\ref{eq:esym6}) of Appendix~\ref{appendix1} by 
taking derivatives of the $G_n (\eta)$ functions defined in Eqs.~(\ref{G1})--(\ref{G6}) and using 
$\displaystyle \frac{\partial G_n (\eta)}{\partial \rho} = \frac{\partial G_n (\eta)}{\partial \eta} \, \frac{\partial \eta}{\partial \rho}$, where $\displaystyle \frac{\partial \eta}{\partial \rho} = \frac{\pi^2 \mu_i}{2 k_F^2}$ for $\eta = \mu_i k_F$.
The results for $\displaystyle \frac{\partial G_n (\eta)}{\partial \eta}$ are given in Eqs.~(\ref{G1'})--(\ref{G6'}). The same procedure can be repeated to compute $\displaystyle \frac{\partial^2 G_n (\eta)}{\partial \rho^2}$. 
As this is relatively straightforward, we omit the explicit results for these derivatives.
\end{widetext}

\mbox{}\newpage

\bibliography{bibtex}
\end{document}